\documentclass[12pt,reqno]{amsart}

\usepackage{amssymb,amsthm,mathtools}
\allowdisplaybreaks

\usepackage{microtype}
\usepackage{enumitem}
\usepackage[a4paper,margin=1.25in]{geometry}
\usepackage{setspace}
\usepackage[T1]{fontenc}
\theoremstyle{plain}
\newtheorem{theorem}{Theorem}

\newtheorem{lemma}{Lemma}
\newtheorem{corollary}{Corollary}

\theoremstyle{definition}
\newtheorem{definition}{Definition}
\newtheorem{example}{Example}
\newtheorem{remark}{Remark}

\usepackage{xcolor}
\usepackage{subcaption}
\usepackage{tikz}
\usepackage{tikz-3dplot}
\usetikzlibrary{patterns,patterns.meta,arrows.meta,calc}
\tikzset{
  every picture/.style={line cap=round, line join=round},
  >=Latex, 
  dot/.style={circle,fill,inner sep=1.1pt}
}

\usepackage{accents}
\usepackage{multicol}
\usepackage{verbatim}

\usepackage{csquotes}
\usepackage[
backend=biber,
style=authoryear-comp,
maxbibnames=9,
maxcitenames=2,
uniquename=false,
uniquelist=false,
natbib,
sorting=nyt,
sortcites=false,
backref=false
]{biblatex}

\usepackage{xurl}                 
\usepackage{hyperref}
\usepackage{bookmark}
\hypersetup{
  pdftitle={Extreme Points in Multi-Dimensional Screening},
  pdfauthor={Patrick Lahr, Axel Niemeyer},
  pdfkeywords={Multi-Dimensional Types, Extreme Points, Rationalizable Mechanisms, Undominated Mechanisms, Multi-Good Monopoly Problem, Delegation, Indecomposable Convex Bodies},
  pdfcreator={LaTeX with hyperref package},
  pdfproducer={LaTeX},
  colorlinks=true,
  linkcolor=[rgb]{0,0,0.6},
  citecolor=[rgb]{0,0,0.6},
  urlcolor=[rgb]{0,0,0.6},
  pdfborder={0 0 0},           
  pdfborderstyle={/S/U/W 0},   
  pdfhighlight=/N,             
  bookmarksnumbered=true,
  pdfdisplaydoctitle=true
}

\usepackage[nameinlink,capitalize,noabbrev]{cleveref}

\makeatletter
\newcounter{examplecont@saved}
\newenvironment{examplecont}[1][]{%
  \setcounter{examplecont@saved}{\value{example}}%
  \edef\examplecont@old{\theexample}%
  \@ifpackageloaded{hyperref}{%
    \ifcsname theHexample\endcsname
      \let\examplecont@restoreH\theHexample
      \edef\examplecont@Hold{\theHexample}%
    \else
      \let\examplecont@restoreH\relax
      \edef\examplecont@Hold{\theexample}%
    \fi
    \def\theHexample{\examplecont@Hold.cont}%
  }{%
    \let\examplecont@restoreH\relax
  }%
  \renewcommand{\theexample}{\examplecont@old\ (continued)}%
  \begin{example}[#1]%
}{%
  \end{example}%
  \setcounter{example}{\value{examplecont@saved}}%
  \@ifpackageloaded{hyperref}{%
    \ifx\examplecont@restoreH\relax\else
      \let\theHexample\examplecont@restoreH
    \fi
  }{}%
}
\makeatother

\crefname{theorem}{Theorem}{Theorems}
\crefname{lemma}{Lemma}{Lemmas}
\crefname{proposition}{Proposition}{Propositions}
\crefname{corollary}{Corollary}{Corollaries}
\crefname{definition}{Definition}{Definitions}
\crefname{example}{Example}{Examples}
\crefname{remark}{Remark}{Remarks}
\crefname{appendix}{Appendix}{Appendices}

\newcommand{\R}{\mathbb{R}}
\newcommand{\Sph}{\mathbb{S}^{d-1}}

\newcommand{\M}{\mathcal{M}}
\newcommand{\inner}[2]{\langle #1, #2 \rangle}
\newcommand{\ubar}[1]{\underaccent{\bar}{#1}}

\DeclareMathOperator*{\argmax}{arg\,max}

\DeclareMathOperator{\cone}{cone}

\DeclareMathOperator{\conv}{conv}

\DeclareMathOperator{\cl}{cl}
\DeclareMathOperator{\bndr}{bndr}
\DeclareMathOperator{\interior}{int}
\DeclareMathOperator{\ext}{ext}

\DeclareMathOperator{\spn}{span}

\DeclareMathOperator{\aff}{aff}
\DeclareMathOperator{\recc}{recc}
\DeclareMathOperator{\exh}{exh}

\DeclareMathOperator{\Hom}{HC}

\DeclareMathOperator{\epi}{epi}

\DeclareMathOperator{\cconv}{\cl\conv}
\DeclareMathOperator{\menu}{menu}
\DeclareMathOperator{\diam}{diam}
\DeclareMathOperator{\und}{und}
\DeclareMathOperator{\dimension}{dim}

\begin{document}

\title[Extreme Points in Multi-Dimensional Screening]{Extreme Points in Multi-Dimensional Screening}

\thanks{We thank 
 Felix Bierbrauer,  
 Peter Caradonna,
 Simone Cerreia-Vioglio, 
 Roberto Corrao,
 Gregorio Curello, 
  Eddie Dekel,
 Laura Doval,
 Piotr Dworczak,
 Mira Frick,
 Alkis Georgiadis-Harris, 
  Yannai Gonczarowski,
 Daniel Gottlieb,
 Michael Greinecker, 
 Nima Haghpanah,
 Marina Halac,
 Deniz Kattwinkel, 
 Bobby Kleinberg,
 Andreas Kleiner,
 Rohit Lamba,
 Elliot Lipnowski, 
 Fabio Maccheroni,
Kristof Madarasz,
 Alejandro Manelli, 
 Jeffrey Mensch, 
 Benny Moldovanu, 
 Roger Myerson, 
 Efe Ok, 
 Tom Palfrey, 
 Martin Pollrich, 
 Luciano Pomatto,
 Justus Preußer, 
 Phil Reny,
 Kota Saito,
 Larry Samuelson, 
 Fedor Sandomirskiy, 
 Mario Schulz, 
 Alex Smolin,
 Ina Taneva, 
 Omer Tamuz,
 Alex Teytelboym,
 Tristan Tomala,
 Aleh Tsyvinsky, 
 Rakesh Vohra,
 Mark Whitmeyer,
 Frank Yang,
 and various seminar audiences for helpful discussions and comments.}

 \author{Patrick Lahr}\thanks{Lahr: ENS Paris-Saclay,  Department of Economics. \textit{Email: \href{mailto:patrick.lahr@ens-paris-saclay.fr}{patrick.lahr@ens-paris-saclay.fr}}} 
\author{Axel Niemeyer}\thanks{Niemeyer: Caltech, Division of the Humanities and Social Sciences. \textit{Email: \href{mailto:niemeyer@caltech.edu}{niemeyer@caltech.edu}}}
	
\date{\today}

\begin{abstract}
We characterize the extreme points of the set of incentive-compatible mechanisms for screening problems with linear utility. Our framework subsumes problems with and without transfers, such as monopoly pricing, principal-optimal bilateral trade and barter exchange, delegation and veto bargaining, or belief elicitation via proper scoring rules. In every problem with one-dimensional types, extreme points admit a tractable description. In every problem with multi-dimensional types, extreme points are dense in a rich subset of incentive-compatible mechanisms, which we call exhaustive mechanisms. Building on these characterizations, we derive parallel conclusions for mechanisms that can be rationalized as (uniquely) optimal under a fixed objective. For example, in the multi-good monopoly problem, mechanisms that uniquely maximize revenue for some type distribution are dense among all incentive-compatible and individually rational mechanisms. The proofs exploit a novel connection between menus of extreme points and indecomposable convex bodies, first studied by \citet{gale1954irreducible}.

	\bigskip
	\noindent
	\textbf{JEL Codes:} D82, D44,  D86
	
	\noindent
	\textbf{Keywords:} Multi-Dimensional Types, Extreme Points, Rationalizable Mechanisms, Undominated Mechanisms, Multi-Good Monopoly Problem, Delegation, Indecomposable Convex Bodies
\end{abstract}

\setcounter{page}{0}     

\maketitle
\thispagestyle{empty}    

\newpage

\section{Introduction}
\label{sec:intro}

Much of the mechanism design literature assumes that agents' preferences  can be described by a single dimension of private information. Under this assumption, the theory has delivered remarkably clean predictions for optimal mechanisms across various applications.  However, private information is often inherently multi-dimensional,  for instance, in allocation problems with multiple heterogeneous goods or collective decision problems with several alternatives. The literature highlights various forms of complexity of optimal mechanisms in multi-dimensional problems, but explicit descriptions of optimal mechanisms
have not been obtained outside of a few special cases.\footnote{For solutions to special cases, see \citet{wilson1993nonlinear,armstrong1996multiproduct,rochet1998ironing,manelli2006bundling,daskalakis2017strong}; for different forms of complexity of optimal mechanisms, see e.g. \citet{manelli2007multidimensional,daskalakis2014complexity,hart2015maximal,hart2019selling}.}

This paper characterizes the structure of optimal mechanisms for the class of multi-dimensional linear screening problems. In these problems, a principal chooses an allocation that affects their own and an agent's utility. 
Both parties' utilities are linear in allocations and depend on the agent's private information, their type. The principal designs a mechanism to maximize their expected utility given a belief about the agent's type. Linear screening covers various problems with and without transfers such as monopoly pricing, principal-optimal bilateral trade and barter exchange, delegation and veto bargaining, or belief elicitation via proper scoring rules.

Our main results characterize the extreme points of the set of incentive-compatible (IC) mechanisms for linear screening problems. Since the principal maximizes a linear functional---their expected utility---over the set of IC mechanisms, an optimal mechanism can always be found among the extreme points. Thus, extreme points form a candidate set to which the principal can restrict their search for an optimal mechanism, regardless of their utility function and  belief.\footnote{Without any restrictions on the principal's utility function, the set of extreme points is essentially the smallest candidate set to which the principal can restrict attention. Straszewicz's theorem says that a dense subset of extreme points is exposed, i.e., the unique maximizer of some continuous linear functional (e.g., \citet{klee1958extremal}). Any continuous linear functional can be generated by choosing an appropriate utility function for the principal (given any fully supported type distribution), so every candidate set necessarily contains the exposed points.} It is in this sense that determining the structure of optimal mechanisms is tantamount to determining the structure of the extreme points.

The main insight is that in every problem with one-dimensional types, the set of extreme points admits a tractable characterization (\Cref{thm:simple}), whereas in every problem with multi-dimensional types, the set of extreme points is virtually as rich as the set of all IC mechanisms (\Cref{thm:complex}).
Specifically, we identify and characterize a rich set of IC mechanisms, which we call exhaustive mechanisms, and show that the extreme points are ($L^1$-)dense in this set whenever types are multi-dimensional. We provide a simple procedure by which an exhaustive mechanism can be perturbed into an extreme point (\Cref{thm:generic_finite}). 

We demonstrate in several applications that our main qualitative insights obtain even if the principal's utility function is fixed and only the principal's belief is considered a free parameter. 
We define an IC mechanism to be weakly (strongly) rationalizable with respect to a given utility function for the principal if there exists a belief for which this mechanism (uniquely) maximizes the principal's expected utility. We characterize weakly and strongly rationalizable mechanisms in sample applications to monopoly pricing, delegation, and veto bargaining. The strongly rationalizable mechanisms, despite being a subset of the extreme (and exposed) points, remain very rich whenever types are multi-dimensional. For example, in the classical multi-good monopoly problem, the set of strongly rationalizable mechanisms is dense in the set of all incentive-compatible and individually rational mechanisms (\Cref{thm:monopoly}).

These results speak to the capabilities and limitations of the classical Bayesian mechanism design paradigm. An important pillar for the success of the theory is that, in many applications, it makes predictions for optimal mechanisms that are independent of the underlying model parameters. In our setting, these parameters are the principal's utility function and the type distribution. Our results confirm that the theory is able to make sharp predictions about optimal mechanisms for one-dimensional linear screening problems. At the same time, our results demonstrate that the theory can rationalize virtually all IC mechanisms as \textit{uniquely} optimal in multi-dimensional linear screening problems. As shown in the applications, this holds even if the principal's utility function is fixed. To the extent that the type distribution is unobservable to an analyst, the Bayesian model yields virtually no testable implications in multi-dimensional problems. And to the extent that a designer may be uncertain about the type distribution, the model gives little normative guidance for how an optimal mechanism should be designed.

The extreme-point approach applied here has seen successful applications in a number of mechanism design settings, but with the exception of \citet{manelli2007multidimensional} (MV), it has not been applied to settings with multi-dimensional types. We significantly extend MV’s analysis along several dimensions. Our explicit characterizations of extreme points, in particular their denseness in the exhaustive mechanisms, and our characterizations of rationalizable mechanisms are novel, as are the conclusions that can be drawn from these results. We provide a detailed comparison with MV and situate our results within the literature in \Cref{sec:literature}.

At the heart of our arguments is a literature in geometry studying extremal elements of spaces of convex sets.\footnote{They are referred to as indecomposable convex bodies and have been introduced by \citet{gale1954irreducible}. See \citet[Chapter 3.2]{schneider2014convex} or \citet[Chapter 6]{Villavicencio2024} for textbook treatments.} Convex sets correspond to menus which the principal might offer to the agent. We prove a strong version of the taxation principle: the spaces of IC mechanisms, convex menus, and indirect utility functions are affinely isomorphic (\Cref{lem:isomorphisms}). This allows us to transfer several fundamental results from the mathematical literature.

\subsection*{Structure of the Paper.}	
\Cref{sec:notation} introduces relevant notation and mathematical definitions.
\Cref{sec:model} introduces the model. 
\Cref{sec:exhaustive} defines and characterizes exhaustive mechanisms, the set of mechanisms in which extreme points are dense in multi-dimensional problems. 
\Cref{sec:main} presents our main results 
about extreme points, along with several supporting results about the structure of extreme points.
\Cref{sec:applications} discusses applications to monopoly pricing, delegation, veto bargaining, and other problems and identifies the rationalizable mechanisms in the three main applications. 
\Cref{sec:proof_sketches} introduces the relevant mathematical tools and sketches the proofs of all results. 
\Cref{sec:literature} discusses the literature in detail.
\Cref{sec:conclusion} concludes.
\Cref{app:exhaustive,app:main,app:applications,app:technical} contain all proofs in chronological order.

\section{Notation and Mathematical Definitions}
\label{sec:notation}

Let $E$ be a normed vector space. For $X\subseteq E$, we denote by $\interior X$, $\bndr X$, $\cl X$, $\conv X$, $\cone X$, and $\aff X$ the interior, boundary, closure, convex hull, conical hull, and affine hull of $X$, respectively. 
Suppose $X\subseteq E$ is convex. $\ext X$ denotes the set of \textbf{extreme points (vertices)} of $X$, i.e.,  those $x\in X$ for which $x=\lambda x' + (1-\lambda) x''$ and $\lambda\in (0,1)$ implies $x=x'=x''$. $\exp X$ denotes the set of \textbf{exposed points} of $X$, i.e.,   those $x\in X$ for which there exists a continuous linear functional $f:E\to\mathbb{R}$ such that $f(x)>f(x')$ for all $x'\in X$, $x\neq x'$.  Every exposed point is extreme, but the converse is not generally true. A \textbf{face} $F$ of $X$ is a convex subset of $X$ such that for all $x\in F$, $x',x''\in X$, and $\lambda\in(0,1)$, $x=\lambda x'+(1-\lambda) x''$ implies $x',x''\in F$. Suppose $E=\mathbb{R}^d$. A \textbf{polyhedron} $P$ is the intersection of finitely many closed halfspaces, a \textbf{polytope} is a bounded polyhedron, and a \textbf{polyhedral cone} is a polyhedron that is also a cone (i.e.,  closed under multiplication with non-negative scalars). A \textbf{facet} $F$ of $P$ is a face of $P$ such that $\dim F=\dim P-1$, where the dimension of a convex set is the dimension of its affine hull. Every subset of Euclidean space is equipped with the Euclidean norm and the Borel sigma-algebra. 

\section{Model and Preliminaries}
\label{sec:model}

\subsection*{Allocations and Preferences.}
There is a principal and an agent. 
The principal chooses an allocation $a\in A\subset\mathbb{R}^d$.
The principal's preferences over allocations depend on the agent's private information, their type $\theta\in\Theta\subset\mathbb{R}^{d}\setminus\{0\}$.
An agent of type $\theta\in\Theta$ derives utility $a\cdot \theta$ from allocation $a\in A$.
Given the agent's type $\theta\in\Theta$, the principal derives utility $a\cdot v(\theta)$ from allocation $a\in A$, where $v:\Theta\to\mathbb{R}^d$ is a bounded measurable function that captures the conflict of interest between both parties.

\subsection*{Mechanisms.}
A direct mechanism is a measurable function $x:\Theta\to A$  that asks the agent to report their type $\theta$ and then implements an allocation $x(\theta)$. By the revelation principle, it is without loss of generality to screen the agent using a direct mechanism that is \textbf{incentive-compatible (IC)}:
\begin{equation}
\label{eq:IC}\tag{IC}
x(\theta)\cdot\theta \;\ge\; x(\theta')\cdot\theta
\quad \forall\, \theta,\theta'\in\Theta .
\end{equation}
IC means that the agent has no incentive to misreport their type. Henceforth, any mechanism is direct.
An \textbf{optimal} mechanism is any solution to the principal's problem
\begin{equation}
	\label{eq:OPT}
	\tag{OPT}
		\sup_{x:\Theta\to A}  \int_\Theta (x(\theta) \cdot v(\theta)) f(\theta) \, d\theta 
		\quad \text{s.t. }    \eqref{eq:IC}, 
\end{equation}
where $f:\Theta\to\mathbb{R}_+$ is a bounded probability density that describes the principal's belief about the agent's type. The density can alternatively be interpreted as the distribution of types in a continuous population.\footnote{Under this interpretation, certain population-level constraints such as budget- or capacity constraints may arise naturally. They can be handled via a result due to \citet{winkler1988extreme}. Winkler's result implies that the extreme points of the set of IC mechanisms subject to $k$ affine constraints are convex combinations of at most $k+1$ extreme points of the unconstrained set of IC mechanisms.}
We can easily incorporate certain types of individual rationality constraints into the analysis; see \Cref{rem:IR} below.

\subsection*{Assumptions.} 
We assume that the allocation space $A$ is a $d$-dimensional polytope, and the set $\{\lambda\theta\mid \theta\in\Theta,\,\lambda\in\mathbb{R}_+\}$ of all rays through the type space $\Theta$ is a $d$-dimensional polyhedral cone.\footnote{The restriction to polyhedral rather than convex sets is for technical convenience. If the dimensions of $A$ and $\cone\Theta$ did not match, then there would either be allocation directions orthogonal to the type space (and hence all types would be indifferent to these directions) or there would be type directions orthogonal to the allocation space (and hence the agent's preferences would not change in these directions). We implicitly prune these orthogonal directions from the model.} We say that the type space is \textbf{unrestricted} if $\cone\Theta=\mathbb{R}^d$.
Note that types on the same ray from the origin have the same preferences over the allocations in $A$. Therefore, we shall select from each ray $\mathbb{R}_+\theta=\{\lambda\theta\mid \lambda\in\mathbb{R}_+\}$ a representative type. Without loss of generality, we henceforth assume one of two normalizations:
    \begin{enumerate}
        \item \textit{non-transferable utility:} $\Theta\subseteq\mathbb{S}^{d-1}=\{y\in\mathbb{R}^d:\: ||y||=1\}$;
        \item \textit{transferable utility:} for each $\theta\in\Theta$, $\theta_d=-1$, i.e.,  the last allocation dimension is interpreted as a numeraire for which the agent has a known marginal utility. It then makes sense to also assume that $v(\theta)=(\ldots,1)$ for all $\theta\in\Theta$, and $A=\tilde A\times[-\kappa_p,\kappa_a]$, where $\kappa_p$ and $\kappa_a$ are the principal's and the agent's endowment of numeraire, respectively.
    \end{enumerate}
	Given normalized types, a $d$-dimensional allocation space $A$  corresponds to a $(d-1)$-dimensional type space $\Theta$.\footnote{Contrary to other notions of one-dimensionality in the mechanism design literature (see, e.g.,  \citet[Chapter 5.6]{borgers2015introduction}), a one-dimensional type space need here not imply a linear order on the underlying preferences. For example, $\Theta=\mathbb{S}^1$ may be a circle.} 
    We equip the type space with the canonical $(d-1)$-dimensional (Hausdorff) measure.

\subsection*{Menus and Payoff-Equivalence.}
By the taxation principle, instead of designing a mechanism, the principal can equivalently
offer the agent a menu (or delegation set) $M\subseteq A$. 
Any choice function  
$
 x(\theta)\in \argmax_{a\in M} a \cdot\theta
 $
 defines an IC mechanism $x:\Theta\to A$. The value function $U(\theta)=\theta\cdot x(\theta)$ is the agent's \textbf{indirect utility function} associated with the mechanism $x$.
Mechanisms defined by the same menu are \textbf{payoff-equivalent}, that is,
the associated indirect utility functions are the same.
For IC mechanisms, payoff-equivalence is equivalent to equality almost everywhere.\footnote{
Payoff-equivalent IC mechanisms are subgradients of the same indirect utility function (\Cref{lem:isomorphisms}). Indirect utility functions are convex, in fact, sublinear on $\cone\Theta$  (convex and 1-homogeneous), hence differentiable almost everywhere. Thus, payoff-equivalent mechanisms are equal almost everywhere.
	 Conversely, knowing almost everywhere any subgradient of a sublinear function suffices to uniquely determine the function itself.} Thus, payoff-equivalent mechanisms also yield the same expected utility to the principal (because the belief is given by a density $f$). 
We identify payoff-equivalent mechanisms, that is, $x=x'$ if $x(\theta)=x'(\theta)$ for almost every $\theta\in\Theta$, and write $\mathcal{X}$ for the set of payoff-equivalence classes of IC mechanisms.\footnote{Alternatively, we could assume a tie-breaking rule for the agent; see an earlier version of this paper, in particular, Figure 6 and the remark thereafter, for further discussion (arXiv:2412.00649v1).}
In \Cref{app:isomorphisms} (\Cref{lem:Hausdorff}), we show that $\mathcal{X}$ is $L^1$-compact. Therefore, a solution to \eqref{eq:OPT} exists and, since $\mathcal{X}$ is convex, can be found among the extreme points $\ext\mathcal{X}$ of $\mathcal{X}$.
We refer to the essential range\footnote{The essential range of a measurable map $x:\Theta\to A$ is the support of the distribution on $A$ induced by $x$:
$$
\menu(x)=\{a\in A\mid x^{-1}(\tilde A)  \text{ has non-zero measure for all open neighborhoods } \tilde A\ni a \}.
$$
Equivalently, 
for the associated indirect utility function $U$,
$
\menu(x)=\cl\{\nabla U(\theta):\: U \text{ is differentiable at } \theta\}.
$
See \Cref{lem:isomorphisms} in \Cref{sec:proof_sketches}.} of a mechanism $x\in\mathcal{X}$ as its \textbf{essential menu} $\menu(x)$. 
Compared to the range $x(\Theta)$, the essential range filters out allocations made only to the measure zero set of indifferent types. 
For instance, if the \textbf{menu size} $|\menu(x)|$ is finite, then the essential menu is the set of allocations that are made by the mechanism with strictly positive probability (cf.\ \citet[Definition 7]{daskalakis2017strong}). 
For brevity, we henceforth omit the qualifier ``essential.''

\begin{remark}[Individual Rationality]
\label{rem:IR}
	In some screening problems, there may be a veto allocation $\ubar a\in A$ that the agent can enforce unilaterally. For these problems, the revelation principle implies that IC mechanisms $x:\Theta\to A$ must  also be \textbf{individually rational (IR)}:
	\begin{equation}
		\label{eq:IR}
		\tag{IR}
		x(\theta)\cdot\theta\geq \ubar a\cdot\theta\quad\forall\theta\in\Theta.
	\end{equation}
	Any characterization of the extreme points of the set of IC mechanisms is valid for IC and IR mechanisms under an additional assumption. Specifically, suppose there exists a type $\ubar\theta\in\Theta$ such that $\ubar a \in\argmax_{a\in A} a\cdot\ubar\theta$ and, for all $\tilde a\in \argmax_{a\in A} a\cdot\ubar\theta$ and all $\theta\in\Theta$, $\tilde a\cdot\theta\geq \ubar a\cdot\theta$. This assumption, which is satisfied in all applications in \Cref{sec:applications}, yields the familiar structure where IR for a specific type, here $\ubar\theta$, implies IR for every type. Under this assumption, the extreme points of the set of IC and IR mechanisms are exactly the extreme points of the set of IC mechanisms that also satisfy IR (in general, the latter is a subset of the former).\footnote{To see this, note that $F=\argmax_{a\in A} a\cdot\ubar\theta$ is a face of $A$. An IC  mechanism $x$ is IR if and only if $\menu(x)\cap F\neq\emptyset$. Since $F$ is a face, if $x=\frac{1}{2} x' + \frac{1}{2}x''$ for IC mechanisms $x'$ and $x''$, then $\menu(x')\cap F$ and $\menu(x'')\cap F$. Hence, the decomposing IC mechanisms $x'$ and $x''$ automatically satisfy IR.
	} 
	Even if the assumption is not met, our qualitative insights about multi-dimensional type spaces remain unchanged since IR constraints only inject additional complexity into the set of extreme points.

\end{remark}

\section{Feasibility Constraints and Exhaustive Mechanisms} 
\label{sec:exhaustive}

 In this section, we identify a simple necessary condition for an IC mechanism to be an extreme point. We later show that this condition, which we call exhaustiveness, is almost sufficient if types are multi-dimensional. Intuitively, exhaustiveness requires that a mechanism leaves no slack in the feasibility constraints that define the allocation space $A$.

    To define the feasibility constraints, recall that the allocation space $A$ is a polytope. Thus, there exists a finite set $\mathcal{F}$ of facet-defining hyperplanes $H=\{y\in\mathbb{R}^d\mid y\cdot n_H = c_H\}$ of $A$. That is, $A=\bigcap\{H_{-}:\:H\in\mathcal{F}\}$, where $H_{-}=\{y\in\mathbb{R}^d\mid y\cdot n_H \leq c_H\}$ are the associated halfspaces containing $A$. Each halfspace corresponds to an affine restriction on the space of available allocations, and no restriction is redundant given the others. For example, a quadrilateral in the plane is described four feasibility constraints, a pentagon by five feasibility constraints, and so forth.
	We denote the set of binding feasibility constraints of mechanism $x$ by  
	\begin{equation*}
		\mathcal{F}(x)=\{H\in\mathcal{F}\mid \menu(x)\cap H\neq\emptyset\}.
	\end{equation*}

\begin{definition}
		\label{def:exhaustive}
        Mechanisms $x,x'\in \mathcal{X}$ are \textbf{homothetic} if there exists $\lambda\in\mathbb{R}_{++}$ and $t\in\mathbb{R}^d$ such that $x=\lambda x'+t$.   
		A mechanism $x\in \mathcal{X}$ is \textbf{exhaustive} 
		if there does not exist another mechanism $x'\in \mathcal{X}$  homothetic to $x$  such that $\mathcal{F}(x)\subseteq\mathcal{F}(x')$. 
	\end{definition}
	
	Two mechanisms are homothetic if one can be obtained from the other by positive scaling and translation. In geometric terms, a homothety preserves the shape and orientation of menus. 
	In economic terms, a homothety preserves the agent's ordinal preferences over menu items and, in particular, the binding incentive constraints. If a mechanism is not exhaustive, that is, further feasibility constraints can be made binding without changing the binding IC constraints, then the mechanism cannot be an extreme point of the set of IC mechanisms:

    \begin{lemma}\label{lem:exhaustive_no_homothetic}
        A mechanism $x\in \mathcal{X}$ is exhaustive if and only if there do not exist distinct mechanisms $x',x''\in\mathcal{X}$ such that 
       $x'$ or $x''$ is homothetic to $x$ and  $
       x=\frac{1}{2}x'+\frac{1}{2}x''
       $.
	\end{lemma}

    We proceed by characterizing the set of exhaustive mechanisms more explicitly. 
    As we illustrate below, this characterization can be used to infer common properties of exhaustive mechanisms and, a fortiori, of extreme points.
    Recall that $n_H$ is the normal vector of the facet-defining hyperplane $H\in\mathcal{F}$ of the allocation space $A$.

\begin{lemma}
		\label{lem:exhaustive_characterization}
		A constant mechanism $x\in \mathcal{X}$ is exhaustive if and only if $\menu(x)=\{a\}$ for $a\in\ext A$.
        A non-constant mechanism $x\in \mathcal{X}$ is exhaustive if and only if (1) $\spn \{n_H\}_{H\in\mathcal{F}(x)} =\mathbb{R}^d$ and (2) $\bigcap\limits_{H\in\mathcal{F}(x)} H=\emptyset$.  
	\end{lemma}

That is, a non-constant mechanism is exhaustive if and only if the facet-defining hyperplanes corresponding to the binding feasibility constraints satisfy two conditions: (1) their normal vectors span the ambient space and (2) they have an empty intersection. Equivalently, $\mathcal{F}(x)$ contains a subset of $d+1$ hyperplanes of which (1) $d$ hyperplanes intersect in a single point $a$ and (2) the remaining hyperplane does not contain $a$. In particular, if $d=2$, then a non-constant mechanism $x\in\mathcal{X}$ is exhaustive if and only if $|\mathcal{F}(x)|\geq 3$.  The two conditions ensure that a mechanism cannot be translated or scaled in a way that makes additional feasibility constraints binding. \Cref{fig:exhaustive_characterization} illustrates.

\begin{figure}
    \centering
    \begin{tikzpicture}[>=Stealth,scale=0.8]
        \coordinate (A2) at (0,0);
        \coordinate (B2) at (5,0);
        \coordinate (C2) at (5,4);
        \coordinate (D2) at (0,4);

        \draw[thick] (A2) -- (B2) -- (C2) -- (D2) -- cycle;

        \coordinate (P2) at (1,0);
        \coordinate (Q2) at (3,4);

        \coordinate (P3) at (3,0);
        \coordinate (Q3) at (5,4);

        \draw[dotted, thick] (P2) -- (Q2) node[pos=0.7, sloped, above left] {$\menu(x)$};
        \draw[dotted, thick] (P3) -- (Q3);

        \foreach \point in {P2,Q2,P3,Q3}
        \filldraw [black] (\point) circle (2pt);

        \draw[->,thick] (P2) -- ($(P2)!0.55!(P3)$);
        \draw[->,thick] (Q2) -- ($(Q2)!0.55!(Q3)$);

    \end{tikzpicture}
    \hspace{1cm}
    \begin{tikzpicture}[>=Stealth,scale=0.8]
        \coordinate (A) at (-1,0);
        \coordinate (B) at (4,0.5);
        \coordinate (C) at (3,4);
        \coordinate (D) at (-1,3);
        \coordinate (E) at (-2,1);

        \coordinate (F) at (-2.58, -0.16);

        \draw[dashed] ($(A)!-0.4!(B)$) -- ($(A)!1.2!(B)$);
        \draw[dashed] ($(D)!-0.5!(E)$) -- ($(D)!1.7!(E)$);

        \filldraw (F) circle (2pt);
        \node at (-2.58, -0.5) {$\bigcap_{H\in\mathcal{F}(x')} H$};

        \draw[thick] (A) -- (B) -- (C) -- (D) -- (E) -- cycle;

        \coordinate (P) at (1,0.2);
        \coordinate (Q) at (-1.5,2);

        \draw[dotted, thick] (P) -- (Q) node[pos=0.975, sloped, below right] {$\menu(x')$};

        \foreach \point in {P,Q}
        \filldraw [black] (\point) circle (2pt);

        \coordinate (FP2) at ($(F)!1.46!(P)$);
        \filldraw (FP2) circle (2pt);

        \coordinate (FQ2) at ($(F)!1.46!(Q)$);
        \filldraw (FQ2) circle (2pt);

        \draw[dotted, thick] (FP2) -- (FQ2);

        \draw[->] (P) -- ($(P)!0.55!(FP2)$);
        \draw[->] (Q) -- ($(Q)!0.65!(FQ2)$);
    \end{tikzpicture}
    \caption{Illustrations of conditions (1) and (2) from \Cref{lem:exhaustive_characterization}. 
    Left: Condition (1) is violated by a menu touching two parallel facets of a rectangle. The menu can be translated horizontally until it touches an inclusion-wise larger set of facets.
    Right: Condition (2) is violated by a menu that touches only two facets of a pentagon. The menu can be scaled relative to the intersection point of the two facet-defining lines until it touches an inclusion-wise larger set of facets.}
    \label{fig:exhaustive_characterization}
\end{figure}

	\begin{example}
 \label{example:delegation}
		Suppose the allocation space $A=\{a\in\mathbb{R}^d_+\mid \sum_{i=1}^d a_i\leq 1\}$ is the unit simplex. The unit simplex is the allocation space when considering lotteries over $d+1$ alternatives or when dividing time or a budget across $d+1$ options (see \Cref{sec:applications} for applications).
		By \Cref{lem:exhaustive_characterization}, a non-constant mechanism $x\in\mathcal{X}$ is exhaustive if and only if it makes all $d+1$ feasibility constraints of the simplex binding. A feasibility constraint is characterized by those lotteries in which some alternative is chosen with probability zero. Therefore, in economic terms, a non-constant exhaustive mechanism must \textit{grant a strike}: for each alternative, there must be an option in $\menu(x)$ where that alternative is chosen with probability zero. A constant mechanism is exhaustive if and only if it \textit{dictates an alternative}: $\menu(x)=\{a\}$ for some $a\in\ext A$.
	\end{example} 
   
    \section{Main Results}
	\label{sec:main}

    In this section, we present our main results about the extreme points of the set of IC mechanisms. The main insight is that the set of extreme points has a simple structure in \textit{every} problem with one-dimensional types but is virtually as rich as the set of exhaustive mechanisms in \textit{every} problem with multi-dimensional types. 
    For the results, recall that a $d$-dimensional allocation space $A$ corresponds to a $(d-1)$-dimensional type space $\Theta$. Also recall that $\mathcal{F}$ is the set of feasibility constraints defining the allocation space.

    \begin{theorem}
		\label{thm:simple}
        Suppose $d=2$. Then, every extreme point $x\in \ext \mathcal{X}$ is exhaustive and satisfies $|\menu(x)|\leq |\mathcal{F}|$. 
	\end{theorem}

     \Cref{thm:simple} is the essential insight of a complete characterization of the extreme points for problems with one-dimensional types (\Cref{thm:mielczarek} in \Cref{app:extreme_1D}):
     extreme points can be succinctly described as choice functions from a limited number of menu items, akin to the well-known posted-price result for the monopoly problem (\citet{myerson1981optimal}; \citet{riley1983optimal}). The bound given in \Cref{thm:simple} is tight for the unrestricted type space and attained by allocating to each type one of their most preferred extreme points of the allocation space $A$. 
     
\begin{examplecont}
Suppose $d=2$ and the allocation space is the unit simplex. By \Cref{thm:simple}, every extreme point has menu size at most three and must either dictate an alternative or grant a strike. The complete characterization (\Cref{thm:mielczarek}) shows that the converse holds if the type space is unrestricted.
\end{examplecont}

In the multi-dimensional case, the structure of extreme points is fundamentally different
and markedly more complex. To make this point, we equip the set of IC mechanisms $\mathcal{X}$ with the $L^1$-norm 
\begin{equation*}
	||x||_1=\int_\Theta ||x(\theta)|| \, d\theta.
\end{equation*}
Intuitively, two mechanisms $x,x'\in\mathcal{X}$ are ``close'' under this norm if the allocations $x(\theta)$ and $x'(\theta)$ are ``close'' under the Euclidean norm for all but a vanishingly small measure of types $\theta
\in\Theta$. 
We say that a property holds for \textbf{most} elements of a subset of a topological space if it holds on a dense set that is also a countable intersection of relatively open sets (i.e.,  a dense $G_\delta$); this is a standard notion of topological genericity.

\begin{theorem}
    \label{thm:complex}
    Suppose $d\geq 3$. Then, every extreme point of the set of IC mechanisms is exhaustive and most exhaustive IC mechanisms are extreme points. 
\end{theorem}

\begin{examplecont}
	Suppose $d\geq 3$ and the allocation space is the unit simplex. By \Cref{thm:complex}, the set of extreme points is dense in the set of mechanisms that dictate an alternative or grant a strike.
\end{examplecont}

A key implication of \Cref{thm:complex} is that, without any restrictions on the principal's objective $v:\Theta\to\mathbb{R}^d$, only few mechanisms can be ruled out as candidates for an optimal mechanism in any given multi-dimensional problem. Specifically, it follows from Straszewicz's theorem (e.g.,  \citet{klee1958extremal}) and the Riesz representation theorem (e.g.,  \citet[Theorem IV.1.1]{DiestelUhl1977}) that, for every exhaustive mechanism $x\in \mathcal{X}$, every belief density $f$ with full support, and every $\varepsilon>0$, there exists a mechanism $x'\in \mathcal{X}$ and an objective $v$ such that $||x-x'||_1<\varepsilon$ and such that $x'$ is \emph{uniquely} optimal given $v$ and $f$. For the applications below (\Cref{sec:applications}), we show that the same conclusion holds, with only minor application-specific adjustments, if instead the objective $v$ is fixed and the belief density $f$ is considered a free parameter.
This conclusion  contrasts sharply with \Cref{thm:simple} and, more generally, with the many parameter-independent predictions that mechanism design theory has delivered for one-dimensional problems. 

Since exhaustiveness is a property of binding feasibility constraints, \Cref{thm:simple,thm:complex} together show that properties of binding incentive constraints further discipline the set of exhaustive mechanisms if and only if the type space is one-dimensional. Thus, our results corroborate the heuristic understanding in the mechanism design literature that the difficulty with multi-dimensional screening lies in identifying the incentive constraints that are binding in an optimal mechanism.

\subsection{Additional Results}

We present more fine-grained results about extreme point mechanisms in multi-dimensional problems. While \Cref{thm:complex} shows that extreme points are dense in the set of exhaustive mechanisms, denseness alone leaves open whether the complexity of the set of extreme points can be curbed by restricting attention to ``simple'' mechanisms.

We first provide a genericity condition, together with a minor additional condition, under which an exhaustive finite-menu mechanism is an extreme point, ruling out the idea that simple extreme points are particularly special. 
For a mechanism $x\in\mathcal{X}$, we say that $\menu(x)$ is \textbf{in general position} if every affine subspace of $\mathbb{R}^d$ of dimension $1\leq k\leq d-1$ intersects $\menu(x)$ in at most $k+1$ points.
We say that $\menu(x)$ is  \textbf{non-separating} if, for every allocation $a\in\menu(x)$, the set $\{\theta\in\Theta\mid a\notin\argmax_{\tilde a\in \menu(x)} \tilde a \cdot \theta \}$ is connected. Non-separation requires that no subset of types that pool on the same option separate the type space into two components.

\begin{theorem}
    \label{thm:generic_finite}
    Suppose $d\geq 3$. If $x\in\mathcal{X}$ is exhaustive and $\menu(x)$ is finite, in general position, and non-separating, then $x\in\ext\mathcal{X}$.
\end{theorem}

If the type space $\Theta$ is unrestricted, that is, $\cone\Theta=\mathbb{R}^d$, it is readily verified that every mechanism $x\in\mathcal{X}$ automatically has a non-separating menu. If the type space is restricted, then non-separation is close to necessary for an exhaustive mechanism to be an extreme point, in a sense made precise in \Cref{lem:general_position_indecomposable} (\Cref{sec:proof_sketches}).

Since generic finite subsets of Euclidean space are in general position, we have:

   \begin{corollary}
   	\label{cor:open_dense_finite}
   	Suppose $d\geq 3$. For every $k\in\mathbb{N}$, within the set of exhaustive mechanisms of menu size $k$ whose menus are non-separating, the subset of extreme points is relatively open and dense.
   \end{corollary}
   
Moreover, one obtains a non-separating menu by splitting menu items that separate the type space into two almost identical copies of itself. Thus:

    \begin{corollary}
    	\label{cor:dense}
        Suppose $d\geq 3$. The set of extreme points of finite menu size is dense in the set of exhaustive mechanisms.
    \end{corollary}

These results demonstrate that the complexity of the set of extreme points cannot be avoided by restricting attention to mechanisms that make only a limited number of allocations. 
Nevertheless, finite-menu extreme points are nowhere near sufficient to describe the full set of extreme points:

\begin{theorem}
		\label{thm:infinite}
		Suppose $d\geq 3$. Most exhaustive mechanisms are extreme points of uncountable menu size. 
	\end{theorem}

    The proof of \Cref{thm:infinite} establishes the stronger claim that most exhaustive mechanisms are continuous functions (for which the menu is a connected subset of the allocation space). While examples of extreme points with uncountable menu size have been documented in the literature (\citet{manelli2007multidimensional}; \citet{daskalakis2017strong}), the existence and prevalence of continuous extreme points is novel.

\section{Applications}
	\label{sec:applications}

In this section, we discuss the implications of our general insights for specific applications. 
We focus on monopoly pricing, delegation, and veto bargaining, and briefly discuss other applications at the end of this section.

In these applications, the principal's objective $v:\Theta\to\mathbb{R}^d$ is given (e.g.,  revenue maximization), and only the principal's belief $f$ about the agent's type is treated as a free parameter. The challenge in transferring our general insights about extreme points to any given application with a fixed objective is that an extreme point of the set of IC mechanisms need not be optimal with respect to that fixed objective for any belief about the agent's type. 

The main takeaway is that the qualitative insights about extreme points continue to hold, with only minor application-specific adjustments, even if one restricts attention to those extreme points that can actually be optimal---even uniquely optimal---for the given objective. To make this point precise, we define and analyze \emph{rationalizable} mechanisms. 

Henceforth, whenever an application features a veto allocation $\ubar a\in A$ for the agent, we refer to $\mathcal{X}$ as the set of (payoff-equivalence classes) of \eqref{eq:IC} and \eqref{eq:IR} mechanisms.

\begin{definition}
For a given objective $v:\Theta\to\mathbb{R}^d$, a mechanism $x \in \mathcal{X}$ is \textbf{weakly rationalizable} if there exists a bounded belief density $f$ such that
$x$ maximizes 
$\int_{\Theta} \left(v(\theta)\cdot y(\theta)\right)\, f(\theta)\, d\theta$
over all $y\in\mathcal{X}$.
It is \textbf{strongly rationalizable} if, in addition, $x$ is the unique maximizer of  $\int_{\Theta} \left(v(\theta)\cdot y(\theta)\right)\, f(\theta)\, d\theta$
 over all $y \in \mathcal{X}$ (up to payoff-equivalence).\footnote{Our results about strong rationalizability hold verbatim if the rationalizing belief density $f$ is required to have full support.}
\end{definition}

Strongly rationalizable mechanisms play a central role in mechanism design. 
A key step in solving any Bayesian mechanism design problem is to reduce the set of all mechanisms to a smaller candidate set among which an optimal mechanism can be found for every possible type distribution. 
Comparative statics with respect to the type distribution are then performed within this set. Any candidate set necessarily contains the strongly rationalizable mechanisms. Conversely, in our applications, an optimal mechanism can always be found in the closure of the strongly rationalizable mechanisms.\footnote{Consider the image of $\mathcal{X}$ under the map $x\mapsto v\cdot x$; this is the set of the induced indirect utility functions of the principal, called $\mathcal{V}$ in \Cref{sec:proof_sketches}. A result due to \citet{lau1976strongly} implies that, for most belief densities $f\in L^\infty$, there is a uniquely optimal indirect utility function in $\mathcal{V}$. Combining \Cref{lem:VtoX} (\Cref{sec:proof_sketches}) with Berge's maximum theorem  yields the claim.} Thus, the strongly rationalizable mechanisms essentially form \textit{the} minimal candidate set. In particular, the common properties of strongly rationalizable mechanisms correspond to the possible parameter-independent predictions about optimal mechanisms.

Such parameter-independent predictions matter for two reasons. First, the designer may be uncertain about the type distribution or may wish to design a mechanism that is effective across a range of different parameters. Second, type distributions may be difficult for an analyst to observe because they may be known only to the entity designing the mechanism or may represent that entity's subjective beliefs; in such cases, the common properties of rationalizable mechanisms amount to the empirically testable implications of the theory.

As an intermediate step in our analysis, we obtain results about weakly rationalizable mechanisms that may be of independent interest. These mechanisms are closely related to undominated mechanisms (\Cref{def:undominated}). However, we find that weak rationalizability is much too permissive, even in one-dimensional problems. Hence, our focus is on strong rationalizability. 

\subsection{Monopoly Pricing}
The monopoly problem, that is, finding revenue-maximizing mechanisms for selling multiple goods, is the most widely studied multi-dimensional screening problem, yet widely regarded as analytically intractable. We establish below for the monopoly problem that the strongly rationalizable mechanisms are dense in the set of all IC and IR mechanisms---a key barrier to obtaining general, explicit solutions to the problem.

	The monopoly problem can be modeled within our setting as follows:
	\begin{itemize}
		\item $A=[0,1]^{m}\times[0,\kappa]$, with $\kappa>m$, where the first $m$ allocation dimensions are the probabilities with which good $i=1,\ldots,m$ is sold to the agent, and the last allocation dimension is the payment from the agent to the principal; 
		\item $\Theta=[0,1]^{m}\times\{-1\}$, i.e.,  the agent has valuations in $[0,1]$ for each good $i=1,\ldots,m$ and money is the numeraire;
		\item $v(\theta)=\bar v=(0,\ldots,0,1)$ for all $\theta\in\Theta$, i.e.,  the principal maximizes expected revenue;\footnote{The standard assumption of zero costs is for simplicity and can be relaxed to constant marginal costs by translating the type space. With decreasing marginal costs, extreme points remain the relevant candidates for optimality (see also the discussion in \citet{manelli2007multidimensional}). With increasing marginal costs, one must instead follow the approach of \citet{rochet1998ironing}.}
        \item $\ubar a=(0,\ldots,0,0)$, i.e.,  the agent can refuse to trade.
	\end{itemize}
   Allocations can also be interpreted as quantities or as multiple attributes of a quality-differentiated product line with a unit-demand agent. The type density $f$ can be interpreted either as the seller's belief about the valuations of a single buyer or as the aggregate demand of a continuum of buyers. 
   
	We say that a mechanism $x\in\mathcal{X}$ \textbf{excludes the lowest type} if $\tilde x(0,\ldots,0,-1)=(0,\ldots,0)$ holds for all IC and IR mechanisms $\tilde x$ that are payoff-equivalent to $x$. We say that a mechanism $x\in\mathcal{X}$ has
	\textbf{no distortion at the top in each dimension} if, for each $\theta\in\Theta$ and $i=1,\ldots,m$, $\theta_i=1$ implies $\tilde x_i(\theta)=1$ for all IC and IR mechanisms $\tilde x$ that are payoff-equivalent to $x$. In other words, every type willing to pay the maximum amount for good $i$ must be allocated good $i$ with probability 1 and must not be indifferent to any option where they are allocated good $i$ with probability less than 1. 

\begin{theorem}
	\label{thm:monopoly}
	The following hold for the monopoly problem:
	
	\begin{enumerate}
		\item Every IC and IR mechanism that excludes the lowest type and has no distortion at the top in each dimension is weakly rationalizable. These mechanisms are dense in the set of all IC and IR mechanisms.
		\item \begin{enumerate}

			\item[(i)] Suppose there is a single good ($m=1$). Every strongly rationalizable mechanism is a posted-price mechanism with price $p \in (0,1)$.\footnote{That is, $x(\theta) = (1,p)$ if $\theta_1 \geq p$ and $x(\theta) = (0,0)$ otherwise.}
			\item[(ii)] Suppose there are multiple goods ($m \geq 2$). Strongly rationalizable mechanisms are dense in the set of all IC and IR mechanisms.
		\end{enumerate}
	\end{enumerate}
\end{theorem}

	\Cref{thm:monopoly} shows that the main qualitative insights about extreme points from \Cref{sec:main} carry over to the monopoly problem. With a single good, it is well known (modulo terminology) that posted-price mechanisms are the only strongly rationalizable mechanisms. The novel result here is that, with multiple goods, every IC and IR mechanism can be perturbed into a mechanism that is uniquely optimal for some belief of the seller. Thus, the model makes sharp predictions for the case of a single good but can strongly rationalize virtually all mechanisms in the case of multiple goods. To the extent that a seller's belief about the value distribution is unobservable, the model can fit essentially any observation. 
	 And to the extent that a seller may entertain several possible beliefs, the model gives little normative guidance on what mechanism a revenue-maximizing seller should use. We note that a broad range of mechanisms is weakly rationalizable even in the one-good case; indeed, Choquet's theorem implies that these are exactly the randomizations over posted-price mechanisms with prices $p \in (0,1)$.\footnote{\citet[Theorem 1]{boergers2025undominated} independently obtain a similar result for the one-good case. We also note that, without requiring absolute continuity of the rationalizing type distribution, it is well known that a truncated Pareto distribution would rationalize any randomization over posted-price mechanisms (and trivially, a point mass at the lowest type $(0,\ldots,0,-1)$ would do as well).
	}

	\subsection{Delegation}
	We apply our results to a screening problem without transfers where a principal relies on the information of a biased agent in order to make an informed decision. Such delegation problems frequently arise in organizational economics and political economy and can be modeled within our setting as follows:
	\begin{itemize}
		\item $A=\{a\in\mathbb{R}^d_+\mid\sum_{i=1}^d a_i\leq 1\}$ is the unit simplex, i.e.,  the allocation space when considering lotteries over $m=d+1$ alternatives, where lotteries can be interpreted as risky courses of action or as budget or time shares;
		\item $\Theta=\mathbb{S}^{d-1}$ is the unrestricted type space, i.e.,  the agent can have all possible von Neumann–Morgenstern preferences over the lotteries in $A$;
		\item there is no veto alternative $\ubar a$ for the agent;
		\item $v(\theta)=\theta-b$, where $b\in\mathbb{R}_{++}$ with $||b||<1$ is the principal's bias relative to the agent (e.g.,  costs of the alternatives that are not internalized by the agent).
	\end{itemize}
	The assumption that $b\in\mathbb{R}_{++}$ simply normalizes the bias so that the origin $a_0=(0,\ldots,0)$ is the \textbf{neutral alternative}, that is, it is preferred by the principal whenever it is preferred by the agent. We say that a lottery $a\in A$ is \textbf{neutral-excluding} if it yields the neutral alternative with probability zero (i.e.,  $\sum_{i=1}^d a_i=1$). 	The assumption $||b||< 1$ means the agent's information can outweigh the principal's bias (since $||\theta||=1$). 
    
	Our delegation problem features multi-dimensional types whenever there are $m\geq 4$ alternatives and thus differs from the classical formulation of delegation problems \`a la \citet{holmstrom1977incentives}, which assume one-dimensional allocation and type spaces and single-peaked (hence non-linear) preferences; see \Cref{sec:literature} for further discussion. Although our formulation of the delegation problem differs from the classical formulation, it captures the same applications while accommodating multi-dimensional preferences.
	
	\begin{theorem}
		\label{thm:delegation}
		The following hold for the delegation problem:
		\begin{enumerate}
			\item Every IC  mechanism whose menu contains the neutral alternative is weakly rationalizable.
			\item \begin{enumerate}
				\item[(i)] Suppose there are three alternatives ($m=3$). Every strongly rationalizable mechanism either dictates the neutral alternative or its menu consists of the neutral alternative, a neutral-excluding lottery, and at most one other item.
				\item[(ii)] Suppose there are four or more alternatives ($m \geq 4$). The strongly rationalizable mechanisms are dense in the set of IC mechanisms whose menu contains the neutral alternative and a neutral-excluding lottery.
			\end{enumerate}
		\end{enumerate}
	\end{theorem}
    
	Thus, the theory makes sharp predictions for problems with three alternatives but rationalizes a vast set of mechanisms once there are four or more alternatives. Three-alternative problems without transfers have previously been studied as the simplest departure from the ubiquitous two-alternative case (e.g.,  \citet{borgers2009efficient});  a complete characterization of optimal mechanisms in a three-alternative problem has, to our knowledge, not been obtained before. Our novel results for the delegation setting suggest a certain tractability of three-alternative problems but also delineate the limits of what can reasonably be expected from optimal Bayesian mechanism design in problems with four or more alternatives, where type spaces become multi-dimensional. It is worth noting that, in contrast to the monopoly problem, not all IC mechanisms can be approximated by strongly rationalizable mechanisms; the inclusion of the neutral alternative and a neutral-excluding lottery are discernible features of strongly rationalizable mechanisms.

\subsection{Veto Bargaining}
As another application, we briefly discuss a variation of the delegation problem where the principal does not rely on the agent's information but must instead seek the agent's approval to implement an alternative over some status quo. Such veto bargaining problems with incomplete information have recently been analyzed by \citet{kartik2021delegation}, who build upon the classical formulation of delegation problems. In our setting, veto bargaining can be modeled as follows:
\begin{itemize}
	\item $A=\{a\in\mathbb{R}^d_+\mid\sum_{i=1}^d a_i\leq 1\}$ is the unit simplex;
	\item $\Theta=\mathbb{S}^{d-1}$ is the unrestricted type space;
	\item $\ubar a =(0,\ldots,0)$ is the agent's veto alternative (e.g.,  a status quo);
	\item $v(\theta)\equiv b$ determines the principal's preferences via the Bernoulli utility vector $b$ independently of the agent's information.\footnote{We could also allow state-dependent preferences, e.g.,  add an outside option to the delegation setting. We then obtain multi-dimensional analogues of models studied in \citet{saran2022dynamic}, \citet{amador2022regulating}, and \citet{kolotilin2019persuasion}. This model specification would yield very similar insights to delegation and veto bargaining regarding rationalizability.}
\end{itemize}
We also assume that (1) $b \in \mathbb{R}_{++}$, that is, the veto alternative is the principal's least preferred alternative, and (2) the principal has a unique favorite alternative $a^*\in A$ (i.e.,  $\argmax_{i} b_i$ is a singleton).

\begin{theorem}
	\label{thm:veto_bargaining}
	The following hold for the veto bargaining problem:
	\begin{enumerate}
		\item Every IC and IR mechanism whose menu contains the veto alternative $\ubar a$ and the principal's most preferred alternative $a^*$ is weakly rationalizable. 
		\item \begin{enumerate}
			\item[(i)] Suppose there are three alternatives ($m=3$). Every strongly rationalizable mechanism has a menu that contains $\ubar a$, $a^*$, and at most one other item.
			\item[(ii)] Suppose there are four or more alternatives ($m \geq 4$). The strongly rationalizable mechanisms are dense in the set of IC and IR mechanisms whose menu contains $\ubar a$ and $a^*$.
		\end{enumerate}
	\end{enumerate}
\end{theorem}

Like the delegation problem, the veto bargaining problem features the simplex as the allocation space and an unrestricted type space, so the extreme points of the set of IC mechanisms in both problems are the same. However, the IR constraint and the state-independent principal utility function imply a slightly different set of rationalizable mechanisms.

\subsection{Further Applications}

\subsection*{Variations and Generalizations of the Monopoly Problem.} Variations include procurement from a supplier with unknown cost, where the principal acts as a buyer, or mixed settings, where the principal is initially endowed with some goods and the agent with the remaining goods. 
Removing transfers yields a barter exchange setting. The unit-demand case of the monopoly problem, as studied for example in \citet{thanassoulis2004haggling} and \citet{pavlov2011optimal}, also fits into our framework by changing the allocation space from a cube to a simplex.

These trading models can be further generalized by endowing the agent with a separate valuation for each bundle $B \subseteq \{1,\ldots,m\}$, which allows for substitutability and complementarity among goods. The allocation space is then the simplex over all $2^m$ bundles, with an additional dimension for transfers. Free disposal, that is, the willingness to pay weakly more for inclusion-wise larger bundles, can be modeled as a family of affine restrictions on the type space. We do not pursue this general model further since the additive case is standard in the literature and since negative results already obtain for the simpler model.

\subsection*{Belief Elicitation.}
As another application, we briefly mention belief elicitation via proper scoring rules (e.g.,  \citet{gneiting2007strictly}). In our framework, belief elicitation can be modeled as follows. Take the type space $\Theta=\{\theta\in\mathbb{R}^d_+\mid \sum_{i=1}^d \theta_i=1\}$ to be the belief simplex over $d$ possible states. Take the allocation space $A=[0,1]^d$ to be the unit cube, where each coordinate specifies the payment to the agent when a particular state realizes, with the maximum payment normalized to 1. Proper scoring rules $x:\Theta\to A$, which incentivize the agent to report their belief truthfully, are equivalent to IC mechanisms.

Recent contributions by \citet{li2022optimization} and  \citet{kleinberg2023u} investigate problems where the principal maximizes a linear functional $L:\mathcal{X}\to\mathbb{R}$ over the set of proper scoring rules.\footnote{In \citet{li2022optimization}, a linear functional arises from maximizing the agent's incentive to acquire information. In \citet{kleinberg2023u}, a linear functional arises from maximizing a certain notion of forecasting regret over the set of proper scoring rules.} Thus, the extreme points of the set of proper scoring rules are the relevant candidates for optimality. When eliciting beliefs about a binary state ($d=2$), the extreme points admit a simple characterization; in particular, if $x\in\mathcal{X}$, then $|\menu(x)|\leq 3$.\footnote{Specifically, it follows from \Cref{thm:mielczarek} that $x\in\ext\mathcal{X}$ if and only if one of the following holds: (1) $\menu(x)$ is an extreme point of $[0,1]^2$; (2) $\menu(x)=\{(1,0),(0,t)\}$ or $\menu(x)=\{(t,0),(0,1)\}$ for $t\in(0,1]$; (3) $\menu(x)=\{(1,0),(0,1),(t_1,t_2)\}$ with $1<t_1+t_2<2$. The reason why an extreme point can have menu size three is analogous to why menu size three may be necessary in  the monopoly problem with a binding price ceiling.} However, if there are at least three states ($d\geq 3$), the complexity of the set of extreme points suggests severe barriers to obtaining any explicit analytic description of optimal scoring rules for any sufficiently rich class of linear functionals that are to be optimized.

\section{Technical Results}
	\label{sec:proof_sketches}

In this section, we explain the most relevant technical details behind our main results and applications. 
Our approach is to translate between extreme points of the set of IC mechanisms and extreme points of the set of all menus. Menus can be identified with convex subsets of the allocation space, allowing us to draw upon a mathematical literature that has characterized extremal---there called indecomposable---elements of spaces of convex sets. 
We also provide new results about undominated mechanisms, which together with our extreme point characterizations allows us to characterize strongly rationalizable mechanisms in the applications in \Cref{sec:applications}.

\subsection{The Equivalence between Mechanisms, Menus, and Indirect Utilities}

We begin by translating from mechanism space to menu space. Recall that $\mathcal{X}$ is the set of all (payoff-equivalence classes of) IC mechanisms $x:\Theta\to A$. 
Let $\mathcal{U}$ denote the set of all indirect utility functions induced by the mechanisms in $\mathcal{X}$; that is,
	$$
	\mathcal{U}=\{U:\theta\mapsto x(\theta)\cdot \theta\mid x\in\mathcal{X}\}.
	$$

We now define the notion of an extended menu. Intuitively, an extended menu is obtained by adding to an existing menu $M\subseteq A$ all allocations that make no type strictly better off relative to $M$. For example, in the monopoly problem, extending a menu means adding all allocations where the agent pays more or gets less.

\begin{definition}
    A closed convex set $M\subset \mathbb{R}^d$ with $\ext M\subset A$ is an \textbf{extended menu} if every point in $\mathbb{R}^d\setminus M$ is strictly preferred by at least one type $\theta\in\Theta$ to every point in $M$. Let $\mathcal{M}$ denote the set of all extended menus.
\end{definition}

If the type space is unrestricted, that is, $\cone\Theta= \mathbb{R}^{d}$, then extended menus are exactly the non-empty compact convex subsets of the allocation space. Otherwise, if the type space is restricted, extended menus are closed convex sets that extend beyond the boundaries of the allocation space, in directions where all types are worse off. Formally, extended menus share the \textbf{polar cone}
$$
\Theta^\circ = \{ y \in \mathbb{R}^d \mid \forall \theta \in \Theta,\ y \cdot \theta \leq 0\},
$$
of the type space $\Theta$ as their common recession cone.\footnote{For a closed convex set $K\subset \mathbb{R}^d$, the \textbf{recession cone} is given by
		$
		\recc(K)=\{y\in \mathbb{R}^d\mid \forall z\in K,\, y+z\in K \}.
		$}
We emphasize that extended menus are a technical tool without physical meaning; the actual menu more realistically corresponds to the set of extreme points of an extended menu.

In the following result, we show that the sets of IC mechanisms, extended menus, and indirect utility functions are isomorphic with respect to their linear and topological structure. To this end, we equip the set $\mathcal{M}$ of extended menus with Minkowski addition and positive scalar multiplication:
\begin{equation*}
	\lambda M + \rho M' = \{\lambda a+\rho a'\mid a\in M, a'\in M'\},
\end{equation*}
where $M,M'\in\mathcal{M}$ and $\lambda,\rho\in\mathbb{R}_+$. We have equipped $\mathcal{X}$ with the $L^1$-norm. We equip $\mathcal{U}$ with the sup-norm and $\mathcal{M}$ with the Hausdorff distance.\footnote{For $U\in\mathcal{U}$ and $M,M'\in\mathcal{M}$, the sup-norm of $U$ is given by $||U||_\infty=\sup_{\theta\in\Theta} |U(\theta)|$ and the Hausdorff distance between $M$ and $M'$ is given by 
$
d(M,M')=\inf\left\{\varepsilon>0:\:
			M\subseteq M'+\varepsilon B  \text{ and } 
			M'\subseteq M+\varepsilon B
			\right\},
$
where $B=\{z\in\mathbb{R}^d:\: ||z||\leq 1\} $ is the unit ball in $\mathbb{R}^d$. The Hausdorff distance is a metric on $\mathcal{M}$ (\Cref{lem:Hausdorff}).}

Recall that a bijection $f:Y\to Z$ between spaces with appropriate linear structure is \textbf{affine} if it commutes with convex combinations, that is, $f(\lambda y+(1-\lambda) y')=\lambda f( y)+(1-\lambda) f(y')$ for all $y,y'\in Y$ and $\lambda\in[0,1]$, and that a bijection $f:Y\to Z$ between topological spaces is a \textbf{homeomorphism} if $f$ and its inverse are continuous. 

\begin{lemma}[Mechanism-Menu-Utility Equivalence]
		\label{lem:isomorphisms}
    The spaces $\mathcal{X}$, $\mathcal{M}$, and $\mathcal{U}$ are affinely homeomorphic. The affine homeomorphisms are given by:
    \begin{itemize}
			\item $\Phi_1:\mathcal{X}\to\M$ where $x\mapsto (\conv \menu(x)+\Theta^\circ)$; 
			\item $\Phi_2:\M\to\mathcal{U}$ where $M\mapsto (\theta \mapsto \sup_{y\in M} y\cdot\theta)$;
			\item $\Phi_3:\mathcal{U}\to\mathcal{X}$ where $U\mapsto \prod_{\theta\in\Theta}(\partial U(\theta)\cap A)$. 
		\end{itemize}
	Moreover, for an IC mechanism $x\in\mathcal{X}$ with associated extended menu $M=\Phi_1(x)\in\M$ and associated indirect utility function $U=\Phi_2(M)\in\mathcal{U}$, $$\menu(x)=\cl\{\nabla U(\theta):\: U \text{ is differentiable at } \theta\}=\cl\ext M.$$
\end{lemma}
That is, $\Phi_1$ maps a (payoff-equivalence class of an)  IC mechanism to the extension of its menu; $\Phi_2$ maps an extended menu to its support function (which is convex and homogeneous of degree 1); $\Phi_3$ maps an indirect utility function to its subdifferential, where all possible selections from the subdifferential define a payoff-equivalence class of IC mechanisms.

\Cref{lem:isomorphisms} is fundamental to our approach because the homeomorphisms between $\mathcal{X}$, $\M$, and $\mathcal{U}$ map extreme points to extreme points.\footnote{Extreme points are usually defined for subsets of vector spaces, which $\mathcal{M}$ is not. However, by \Cref{lem:isomorphisms}, $\mathcal{M}$ can be embedded into a vector space, which justifies our use of the term \textit{extreme point}.} 
Although it is well-known that screening problems can be approached by investigating indirect utility functions (\citet{rochet1987necessary}) or menus (the ``taxation principle'' due to \citet{hammond1979straightforward}), we are not aware of any existing result in the economic literature that makes the equivalence precise to the extent that we do here.

\subsection{Indecomposability and Exhaustiveness} 
	We next explain how extreme points of the set of extended menus $\mathcal{M}$ can be understood via the notion of indecomposability from the mathematical literature and the notion of exhaustiveness defined in \Cref{sec:exhaustive}. 
	
    For extended menus, we say that $M,M'\in\mathcal{M}$ are \textbf{homothetic} if there exists $\lambda\in\mathbb{R}_{++}$ and $t\in\mathbb{R}^d$ such that $M=\lambda M'+t$.
    $M\in\mathcal{M}$ is
     \textbf{exhaustive} if  there is no extended menu $M'\in\mathcal{M}$ homothetic to $M$ such that $\mathcal{F}(M)\subseteq\mathcal{F}(M')$, where 
    $
    \mathcal{F}(M)=\{H\in\mathcal
    F\mid H\cap \ext M\neq \emptyset\}.
    $ 
    Equivalently, $M\in\mathcal{M}$ is exhaustive if and only if
    the associated mechanism $x\in\mathcal{X}$ is exhaustive.  
    
    By definition, $M\in\ext\mathcal{M}$ if and only if there is no $M',M''\in\mathcal{M}$, $M'\neq M''$, such that $M=\frac{1}{2} M'+\frac{1}{2} M''$, and it is illuminating to distinguish the cases where
    \begin{enumerate}
        \item $M'$ or $M''$ is homothetic to $M$; 
        \item $M'$ and $M''$ are not homothetic to $M$.
    \end{enumerate}
    It is immediate from \Cref{lem:exhaustive_no_homothetic,lem:isomorphisms} that $M$ cannot be written as a convex combination of the first type if and only if $M$ is exhaustive. 

    Convex combinations of the second type are closely related to the notion of indecomposability from the mathematical literature. Let $\mathcal{K}$ denote the set of all closed convex sets $K\subset \mathbb{R}^d$ such that every point in $\mathbb{R}^d\setminus K$ is strictly preferred by at least one type $\theta\in\Theta$ to every point in $K$. Equivalently, $\mathcal{K}$ is the set of all closed convex sets in $\mathbb{R}^d$ with recession cone $\recc K=\Theta^\circ$. $\mathcal{K}$ differs from $\mathcal{M}$ in that we do not require $\ext K\subset A$. 

    \begin{definition}
        A closed convex set $K\in\mathcal{K}$ is \textbf{indecomposable} if there do not exist $K',K''\in\mathcal{K}$ such that $K=\frac{1}{2} K'+\frac{1}{2} K''$ and $K'$ and $K''$ are not homothetic to $K$.\footnote{The original definition due to \citet{gale1954irreducible} applies to compact convex sets. The extension to closed convex sets with a common recession cone is due to \citet{smilansky1987decomposability}.}
    \end{definition}
	
	Indecomposability can be interpreted in terms of binding IC constraints. Specifically, an extended menu associated with a finite-menu mechanism is indecomposable if and only if it is uniquely determined,  up to homothetic transformations, from its normal fan, that is, from knowing for any pair of types whether they have the same favorite allocation(s) in the menu (and hence whether they would mimic one another), but without knowing what these allocations are. (The details for this interpretation can be found in an earlier version of this paper (arXiv:2412.00649v1).) Equivalently, indecomposability ensures that no further IC constraints can be made binding, whereas exhaustiveness ensures the same for the feasibility constraints. 
	Put together, we have:

    \begin{lemma}
    \label{lem:indecomposable+exhaustive}
        If an extended menu $M\in\mathcal{M}$ is indecomposable and exhaustive, then $M\in\ext\mathcal{M}$. The converse holds if $A$ is a simplex and $\Theta$ is unrestricted.
    \end{lemma}
    
	Indecomposability is not generally necessary for a mechanism to be an extreme point. Intuitively, this is because binding incentive and feasibility constraints jointly determine whether or not a mechanism is an extreme point. However, for the special case where the allocation space is a simplex and the type space is unrestricted, the role of incentive and feasibility constraints can be separated, and indecomposability becomes necessary.   

    \subsection{Characterizing Indecomposability}
Known results about indecomposability allow us to obtain our main results about extreme points in \Cref{sec:main}. We begin with the planar case corresponding to one-dimensional types, which requires only a minor extension of a result due to \citet{meyer1972decomposing} and \citet{silverman1973decomposition}.
\begin{lemma} 
	\label{lem:plane_indecomposable} 
	Suppose $d=2$. An extended menu $M \in \mathcal{M}$ is indecomposable if and only if $|\ext M|\leq 3$ when $\Theta$ is unrestricted, and $|\ext M|\leq 2$ when $\Theta$ is restricted.
\end{lemma}

The reason that the bound in \Cref{thm:simple} can be greater than 3 is that indecomposability does not yet account for feasibility constraints. Note, however, that in the recurrent \Cref{example:delegation} and the applications to delegation and veto bargaining, where the allocation space is a simplex and the type space is unrestricted, 	\Cref{lem:plane_indecomposable,lem:indecomposable+exhaustive} already yield a complete characterization of the extreme points for $d=2$. The complete characterization for $d=2$ across all allocation and types spaces is more involved (\Cref{thm:mielczarek}). 

The higher-dimensional case also admits a complete characterization for generic finite menus.

\begin{lemma}	\label{lem:general_position_indecomposable}
    Let $M\in\mathcal{M}$ is such that $\ext M$ is finite and in general position. Then, $M$ is indecomposable if and only if $\ext M$ is non-separating. 
\end{lemma}

For the proof, observe first that if $\ext M$ is finite, then $M$ is a polyhedron by the Minkowski-Weyl theorem since it can be written as the convex hull of its extreme points plus the polar of the type space $\Theta^\circ$, which is a polyhedral cone. If an at least three-dimensional polyhedron $M$ has its vertices $\ext M$ in general position, then each of its bounded 2-dimensional faces is a triangle since every plane intersects $\ext M$ in at most three points (in fact, each bounded face is a simplex). By \Cref{lem:plane_indecomposable}, each triangle is indecomposable by itself. If $\ext M$ is also non-separating, then any two triangles can be connected through a sequence of triangles such that consecutive triangles share an edge. This condition ensures that the only polyhedra that decompose $M$ are homothetic to $M$, hence $M$ is indecomposable (see \citet[Theorem 5.1]{smilansky1987decomposability}; \citet[Theorem 6.2.22]{Villavicencio2024}). From \Cref{lem:general_position_indecomposable}, we can immediately deduce \Cref{thm:generic_finite}. The denseness results in \Cref{sec:main} follow since every menu of an exhaustive mechanism can be perturbed into a non-separating menu that is also in general position by splitting the separating menu items and moving the resulting menu into general position, all while preserving exhaustiveness.

For polyhedra with extreme points that are not in general position, non-separation remains necessary for indecomposability but is no longer sufficient. An \textit{explicit} geometric characterization of indecomposability for general polytopes or polyhedra, let alone for arbitrary closed convex sets, has not been obtained in the mathematical literature and is not to be expected.\footnote{\citet[p.~166]{schneider2014convex} writes: ``Most convex bodies in $\mathbb{R}^d$, $d \geq 3$, are smooth, strictly convex and indecomposable. It appears that no concrete example of such a body is explicitly known. This is not too surprising, since it is hard to imagine how such a body should be described.''} However, various sufficient conditions for indecomposability have been identified (see \citet[Chapter 3.2]{schneider2014convex} and \citet[Chapter 6]{Villavicencio2024}). In addition, there exist algebraic characterizations of indecomposability that turn out to generalize the main result in \citet{manelli2007multidimensional} beyond the monopoly problem (see an earlier version of this paper (arXiv:2412.00649v1)). Since menus that are not in general position carry no special economic significance, and since algebraic characterizations yield no substantive economic insight beyond the explicit characterizations we have given already, we shall not pursue these directions further.

\subsection{Dominance and Weak Rationalizability}
Rationalizable mechanisms, which we characterize in the applications in \Cref{sec:applications}, are closely connected to undominated mechanisms. 
 
\begin{definition}
	\label{def:undominated}
	A mechanism $x\in \mathcal{X}$ is \textbf{dominated} by another mechanism $x'\in \mathcal{X}$ if $x'(\theta)\cdot v(\theta)\geq x(\theta)\cdot v(\theta)$ for almost all $\theta\in\Theta$, with strict inequality on a set of types of positive measure.
	A mechanism $x\in \mathcal{X}$ is \textbf{undominated} if it is not dominated by any other mechanism $x'\in \mathcal{X}$.
\end{definition}

 \citet{manelli2007multidimensional,manelli2012multidimensional} show for the monopoly problem that every undominated mechanism is optimal for some belief about the agent's type. This result can be extended from revenue maximization in the monopoly problem to arbitrary objectives and screening problems:

 \begin{lemma}
\label{lem:undominated_optimal}
	Every undominated IC mechanism is weakly rationalizable.
\end{lemma}

Conversely, every mechanism that is optimal for some belief $f$ with full support on $\Theta$ must clearly be undominated. 
To obtain our results about weakly rationalizable mechanisms in \Cref{sec:applications}, we either characterize undominated mechanisms or provide fairly tight sufficient conditions for a mechanism to be undominated (\Cref{lem:undomi_monopoly,lem:undomi_delegation,lem:undomi_veto}). 

The basic idea behind our results about undominated mechanisms across all applications is as follows. If $x'\in\mathcal{X}$ dominates $x\in\mathcal{X}$, then for the associated indirect utility functions $U',U\in\mathcal{U}$ we have
$$
\nabla_{v(\theta)} U'(\theta)\geq \nabla_{v(\theta)} U(\theta)
$$
for almost every $\theta\in\Theta$, where $\nabla_{v(\theta)}$ denotes the directional derivative in direction $v(\theta)$. This is because $x'\in\partial U'$ and $x\in\partial U$ by \Cref{lem:isomorphisms}. Therefore, $U'-U$ is non-decreasing along any integral curve\footnote{An integral curve is a differentiable path $\gamma:[0,1]\to\Theta$ such that $\gamma'(t)=v(\gamma(t))$.}  of the vector field $\theta\mapsto v(\theta)$ (modulo non-differentiabilities addressed in the formal proofs).
In the applications, dominance implies the existence of types that receive the same allocation---hence the same indirect utility---across all undominated mechanisms. For example, in the monopoly problem, every undominated mechanism necessarily excludes the lowest type and allocates the grand bundle of all goods to the highest type. Suppose all integral curves meet the boundaries of the type space in the same types and these types receive the same utility under both $x$ and $x'$. Then, from the monotonicity of $U'-U$ along the curves, we can conclude $U=U'$ and hence $x=x'$, contradicting the initial assumption that $x'$ dominates $x$. We apply variations of this argument in all applications.

Undominated mechanisms are of independent interest, beyond as a tool for understanding Bayesian mechanism design, as dominance is a basic optimality criterion across risk and ambiguity attitudes of the principal.\footnote{The criterion goes back to at least \citet{myerson1983mechanism}. See \citet{boergers2025undominated} for an insightful discussion.} However, on the negative side, our results demonstrate that only relatively few mechanisms can be ruled out using the dominance criterion.

\subsection{Strong Rationalizability}
The final step is to combine our results about weak rationalizability with our results about extreme points to derive implications for strong rationalizability. In the multi-dimensional case, the two sets of results imply that the weakly rationalizable extreme points are dense in the weakly rationalizable exhaustive mechanisms.\footnote{Weak rationalizability implies exhaustiveness only in some applications; in other applications, exhaustiveness imposes additional restrictions. Note that the denseness of extreme points within the weakly rationalizable exhaustive mechanisms does not follow from the denseness of extreme points in the exhaustive mechanisms because the set of weakly rationalizable mechanisms has empty relative interior. In other words, denseness alone only guarantees that every weakly rationalizable exhaustive mechanism can be perturbed into an extreme point but not that this extreme point is itself weakly rationalizable. To conclude the latter, a characterization of both extreme points and weakly rationalizable mechanisms is needed.} It remains to show that strongly rationalizable mechanisms are themselves dense in the weakly rationalizable extreme points. To this end, we modify Straszewicz's theorem, which asserts that the exposed points of a compact convex set are dense in its extreme points, to suit our particular needs.

Let
$$\mathcal{V}=\{\theta\mapsto x(\theta)\cdot v(\theta)\mid x\in \mathcal{X}\}$$
denote the set of the principal's utility functions induced by the set of \eqref{eq:IC} mechanisms. This set of functions $\Theta\to\mathbb{R}$ is $L^1$-compact and convex because it is a continuous image of the $L^1$-compact convex set $\mathcal{X}$ (\Cref{lem:Hausdorff}). Hence, like $\mathcal{X}$, one can study the extreme points of $\mathcal{V}$. The advantage of working with $\mathcal{V}$ rather than $\mathcal{X}$ is that  $\mathcal{V}$ is naturally paired with the space of all bounded functions, that is, signed densities, equipped with the $L^\infty$-norm.

We say that a principal utility function $V\in\mathcal{V}$ has a certain property if it is induced by a mechanism $x\in\mathcal{X}$ with that property.

\begin{lemma}
	\label{lem:v_exposed}
	The strongly rationalizable principal utility functions in $\mathcal{V}$ are dense in the undominated  principal utility functions in $\ext\mathcal{V}$.
\end{lemma}

We can complete the proofs by translating back from $\mathcal{V}$ to $\mathcal{X}$: 

\begin{lemma}
    \label{lem:VtoX}
    Consider the monopoly problem, delegation problem, or veto bargaining problem. If $x,x'\in\mathcal{X}$ are undominated mechanisms such that $x(\theta)\cdot v(\theta)=x'(\theta)\cdot v(\theta)$ for almost every type $\theta\in\Theta$, then $x=x'$.
\end{lemma}

For example, in the monopoly problem, the transfer rule of an undominated mechanism uniquely determines the entire mechanism. 

\section{Related Literature}
\label{sec:literature}

    \citet{manelli2007multidimensional} (MV) is the first, and prior to this paper, only application of the extreme-point approach to multi-dimensional screening. 
    MV provide an algebraic characterization of extreme points for the multi-good monopoly problem (their Theorem 24).\footnote{They show that a mechanism is an extreme point if and only if a certain linear system associated with the mechanism has a unique solution.} From this characterization, MV deduce that extreme points are dense within the subset of mechanisms that (1) have menu size at most the number of goods plus one and (2) feature an exclusion region and no distortion at the top (Remark 25). MV also show that undominated mechanisms are weakly rationalizable in our sense (Theorem 9 and \citet{manelli2012multidimensional}).  

    Our analysis significantly extends theirs along three dimensions. First, our results apply not only to monopoly pricing but to all linear screening problems. Second, we provide explicit extreme point characterizations and establish the denseness of the extreme points in multi-dimensional settings without any restriction on menu size. Third, we obtain parallel conclusions for extreme points that are uniquely optimal for a given principal utility function. Thus, our results reveal a fundamental limitation of the Bayesian mechanism design framework: minor restrictions aside, the theory is not able to make parameter-free predictions about optimal mechanisms in a wide class of multi-dimensional problems. 

        The rest of the literature deploying the extreme-point approach to mechanism design provides characterizations that apply to problems with one-dimensional types.  
        \citet{manelli2010bayesian} use extreme points to establish the equivalence of Bayesian and dominant strategy incentive-compatibility in standard auctions. 
		\citet{kleiner2021extreme} characterize extreme points of certain majorization sets, show how these majorization sets naturally arise as feasible sets in many economic design problems, and consequently obtain various novel insights into these problems.
        \citet{nikzad2022constrained,nikzad2024multi} build on the majorization approach, allowing for additional constraints on the majorization sets. \citet{yang2024monotone} characterize extreme points of sets of distributions characterized by first-order stochastic dominance constraints and apply these results to a range of problems in economic design and other areas.
        \citet{kleiner2024extreme} characterize certain extreme points of the set of measures that are dominated in the convex order by a given measure; their results apply to multi-dimensional Bayesian persuasion but have no obvious applications to mechanism design. \citet{yang2025multidimensional} characterize extreme points of real-valued multivariate monotone functions and their one-dimensional marginals, with various applications to one-dimensional mechanism design; multi-dimensional problems, however, need characterizations of the extreme points of certain sets of (cyclically) monotone vector fields or, equivalently, of monotone \textit{convex} functions, which we implicitly provide in this paper.

The literature on multi-dimensional screening and mechanism design is too large to summarize here. We point to a survey by \citet{rochet2003economics} for work up to the  2000s. Recent work focuses primarily on the multi-good monopoly problem and can be classified into several strands: 
finding conditions for the optimality of common mechanisms such as separate sales or bundling;\footnote{
	\citet{mcafee1989multiproduct},
	\citet{manelli2006bundling},
	\citet{fang2006bundle},
	\citet{pavlov2011optimal},
	\citet{daskalakis2017strong}, 
	\citet{menicucci2015optimality},
	\citet{bergemann2021optimality}, \citet{haghpanah2021pure},
	\citet{ghili2023characterization}, \citet{yang2023nested}.}
establishing duality results that can be used to certify the optimality of a given mechanism;\footnote{
	\citet{daskalakis2017strong},
	\citet{kleiner2019strong},
	\citet{cai2019duality},
	\citet{kolesnikov2022beckmann},
	\citet{kleiner2022optimal}.
}
studying specific structural properties of optimal mechanisms and showing when such structure arises;\footnote{
\citet{mcafee1989multiproduct},
\citet{thanassoulis2004haggling},
\citet{manelli2006bundling}, 
\citet{hart2015maximal},
\citet{babaioff2018optimal},
\citet{ben2022monotonic},
\citet{bikhchandani2022selling}.
}
quantifying the worst-case performance (approximation ratio) of common mechanisms or classes of mechanisms;\footnote{
\citet{hart2017approximate,hart2019selling},
\citet{li2013revenue}, 
\citet{babaioff2017menu},
\citet{rubinstein2018simple},
\citet{hart2019better},
\citet{babaioff2020simple},
\citet{ben2022monotonic}.
}
identifying optimal robust mechanisms under Knightian uncertainty over the set of type distributions;\footnote{\citet{carroll2017robustness}, \citet{deb2023multi}, \citet{che2023robustly}.}
analyzing asymptotics with respect to the number of goods or the seller's information.\footnote{\citet{armstrong1999price}, \citet{bakos1999bundling}, \citet{frick_iijima_ishii_forthcoming}.}
In general, little is known about optimal multi-dimensional mechanism design with multiple agents;\footnote{ \citet{palfrey1983bundling}, \citet{jehiel1999multidimensional},
	\citet{chakraborty1999bundling},
	\citet{jehiel2007mixed}, 
	\citet{kolesnikov2022beckmann}.} see \Cref{sec:conclusion} for a discussion of potential implications of our results for multi-agent settings.

Our paper is orthogonal to the recent developments in the literature on multi-dimensional screening. We do not focus on specific properties of mechanisms or certain classes of mechanisms or attempt to avoid intractabilities by modifying the problem. Instead, we shed light on where these intractabilities originate and identify the limits of the qualitative predictions that can be drawn within the standard Bayesian framework. 
In this sense, our analysis provides a formal rationale for why strong restrictions on parameters or alternative modeling approaches are needed to make progress on multi-dimensional screening. The unified perspective on problems with and without transfers is a distinguishing feature of our approach.

Besides the implications for optimal mechanism design, our extreme point characterizations can also be viewed as results about implementability with multi-dimensional types (cf. \citet{rochet1987necessary}, \citet{saks2005weak}, and \citet{bikhchandani2006weak}) since, by Choquet's theorem, every non-extreme IC mechanism can be represented as a mixture over extreme points.

Our delegation and veto bargaining applications differ from the classical models based on \citet{holmstrom1977incentives,holmstrom1984theory} in two ways:\footnote{See \citet{melumad1991communication}, \citet{martimort2006continuity}, \citet{alonso2008optimal}, \citet{amador2013theory}, 
\citet{kleiner2021extreme},
and \citet{kolotilin2019persuasion} for delegation and \citet{kartik2021delegation}, \citet{amador2022regulating}, \citet{saran2022dynamic}, and \citet{kolotilin2019persuasion} for veto bargaining.} allocations are lotteries over a finite set of alternatives and the agent has arbitrary vNM preferences. Like us, \citet{kleiner2022optimal} studies optimal Bayesian mechanism design in a delegation problem with multi-dimensional type and allocation spaces and affine utilities. 
Kleiner's duality-based approach is complementary to our extreme-point approach. 
\citet{frankel2016delegating} studies multi-dimensional delegation with quadratic utilities in each dimension and i.i.d. types. Assuming  normally distributed types, Frankel shows that delegation to halfspaces is optimal and, for arbitrary distributions, approximately optimal when the number of dimensions is large. \citet{frankel2014aligned} studies this problem using a robust design approach.

Finally, we draw upon the mathematical literature on indecomposable convex bodies. 
\citet{gale1954irreducible} introduced the concept and announced a number of results without proof. 
Gale's results have since been proven, and the literature has provided many results that go beyond Gale's original presentation; see \citet[Chapter 3.2]{schneider2014convex} and \citet[Chapter 6]{Villavicencio2024} for textbook treatments.
In particular, we use results due to \citet{smilansky1987decomposability} about indecomposable polyhedra.
Related results characterize extremal convex bodies within a given compact convex set in the plane (\citet{mielczarek1998extreme, grzaslewicz1984extreme}); see \Cref{thm:simple,thm:mielczarek} for the application in our paper. 
Characterizations of indecomposable convex bodies
can alternatively be seen, via  support function duality, as characterizations of the extremal rays of the cone of sublinear (i.e., convex and 1-homogeneous) functions. A subset of the results known in the literature on indecomposable convex bodies have been independently obtained in studies of the extremal rays of the cone of convex functions by 
\citet{johansen1974extremal} (for two-dimensional domains) and \citet{bronshtein1978extremal} (for $d$-dimensional domains).\footnote{We thank Andreas Kleiner for pointing us to these references.}

		\section{Conclusion}
	\label{sec:conclusion}
	
	We have characterized extreme points of the set of incentive-compatible (IC) mechanisms for linear screening problems. For every problem with one-dimensional types, extreme points admit a tractable description and a tight upper bound on their menu size. In contrast, for every problem with multi-dimensional types, we have identified a large set of IC mechanisms---exhaustive mechanisms---in which the extreme points are dense. 
    We have obtained parallel conclusions about mechanisms that can be rationalized as uniquely optimal in sample applications to monopoly pricing, delegation, and veto bargaining, where the principal's utility function is fixed.

    While one-dimensional problems allow us to make predictions that are independent of the underlying model parameters, such predictions are largely unattainable in multi-dimensional problems. To the extent that the designer may be uncertain about model parameters such as type distributions, the theory does not provide tangible practical guidance for how an optimal mechanism should be designed. And to the extent that these parameters are difficult to observe for an outside analyst, the theory yields virtually no testable implications. In this sense, our analysis provides a formal rationale for why strong restrictions on parameters or alternative modeling approaches are needed to make progress on multi-dimensional screening. 
    
	While our focus has been on screening problems, where there is only a single (representative) agent, implications of our results for multi-agent settings are to be expected. In multi-agent settings, Bayesian incentive compatibility of a given multi-agent mechanism is equivalent to separately requiring incentive compatibility with respect to each agent's interim-expected mechanism (see, e.g.,  \citet[Chapter 6]{borgers2015introduction}). These interim-expected mechanisms, one for each agent, must then be linked to an ex-post feasible mechanism via an appropriate analogue of the Maskin-Riley-Matthews-Border conditions.\footnote{\citet{maskin1984optimal}, \citet{matthews1984implementability}, \citet{border1991implementation}. Recent treatments include \citet{che2013generalized}, \citet{gopalan2018public}, and \citet{ValenzuelaStookey2023Interim}; see these papers for  limitations of the approach.} 
	Thus, if the extreme points in a multi-agent problem were simpler than the extreme points characterized here for the one-agent case, then this reduction in complexity would have to come from these additional feasibility conditions. This is not the case for problems with one-dimensional types (see, e.g.,  \citet{kleiner2021extreme}) and is not to be expected for problems with multi-dimensional types.
	
	Our main methodological contribution is to link extreme points of the set of incentive-compatible mechanisms to indecomposable convex bodies studied in convex geometry. This methodology, where we study incentive-compatible mechanisms by analyzing the space of all menus from which the agent could choose, is potentially useful in other areas of economic theory.
    Examples that come to mind are menu choice \`{a} la \citet{dekel2001representing} and the random expected utility (REU) model of \citet{gul2006random} (to which the linear screening model is closely related). We believe that our characterizations will also prove useful for understanding the reverse question to our approach: For which model parameters is a given extreme point mechanism or class of extreme point mechanisms optimal? We leave this question to future work.

\begin{appendix}

\section{Proofs}
\label{app:full}

\Cref{app:exhaustive,app:main,app:applications,app:technical} collect the proofs for \Cref{sec:exhaustive,sec:main,sec:applications,sec:proof_sketches} in chronological order.
\Cref{app:exhaustive} proves the two characterizations of exhaustiveness. \Cref{app:main} provides the full characterization of extreme points for one-dimensional types omitted from the main text and the proofs of all results about extreme points from the main text. \Cref{app:applications} collects the proofs for the applications to monopoly, delegation, and veto bargaining---in particular, the characterizations of undominated mechanisms. \Cref{app:technical} proves the technical lemmas discussed in \Cref{sec:proof_sketches}, which are used throughout all proofs.

\subsection{Proofs for \texorpdfstring{\Cref{sec:exhaustive}}{Section~\ref{sec:exhaustive}}}
\label{app:exhaustive}

\begin{proof}[Proof of \Cref{lem:exhaustive_no_homothetic}]	
	 
	 As an auxiliary step, we show that the homothety class 
	 $$
	 \Hom(M)=\{(\lambda,t)\in \mathbb{R}_+\times\mathbb{R}^d \mid \lambda \ext M + t\subset A \}
	 $$
	 of $M\in\M$
	 is a polytope.  $\Hom(M)$ is bounded because $A$ is bounded. 
	 Therefore, it suffices to show that $\Hom(M)$ is the intersection of finitely many halfspaces. Define
	 	\begin{align*}
	 	\Hom_-(M,H)&= \{(\lambda,t)\in \mathbb{R}\times\mathbb{R}^d \mid \lambda\ext M + t\subset H_- \}\\
	 	&=  \left\{(\lambda,t)\in  \mathbb{R}\times\mathbb{R}^d \ \middle\vert \ \lambda \max_{a\in \ext M} a \cdot n_H + t \cdot n_H \leq c_H\right\},
	 \end{align*}
	 where $H_-=\{z \in \mathbb{R}^d: z \cdot n_H \leq c_H \}$ is the halfspace that contains $A$ and is bounded by the facet-defining hyperplane $H\in\mathcal{F}$ of $A$. Each $\Hom_-(M,F)$ is a halfspace in $\mathbb{R}^{d+1}$ (with normal $(\max_{a\in \ext M} a \cdot n_H,n_H)$). Thus, letting $Z=\mathbb{R}_{+}\times\mathbb{R}^d$, we conclude that
	 $
	 \Hom(M)=Z\cap \bigcap_{H\in\mathcal{F}} \Hom_-(M,H)
	 $
	 is a polytope.
	 
	 Note that $(\lambda,t)=(1,0)\in\ext\Hom(M)$ if and only if there do not exist $M',M''\in\mathcal{M}$ homothetic to $M$ such that $M=\frac{1}{2}M'+\frac{1}{2}M''$. If $M'$ is homothetic to $M$, then $M''$ is also homothetic to $M$ or $M''=t+\Theta^\circ$ for some $t\in\mathbb{R}^d$.
	 
	 Thus, we complete the proof by showing that $M$ is exhaustive if and only if $(\lambda,t)=(1,0)\in\ext\Hom(M)$. 
	 Note that $(\lambda,t)=(1,0)$ does not lie on the boundary of $Z$.
	 Every other halfspace $\Hom_-(M,H)$ of $\Hom(M)$ corresponds to a facet-defining hyperplane $H\in\mathcal{F}$ of $A$, and $H\in\mathcal{F}(\lambda M+t)$ if and only if $(\lambda,t)\in\bndr \Hom_{-}(M,H)$ (i.e.,  $\lambda\max_{a\in \ext M} a\cdot n_H+t\cdot n_H=c_H$). By definition, $M$ is exhaustive if there is no $(\lambda,t)\in \Hom(M)$ such that $\mathcal{F}(M)\subseteq\mathcal{F}(\lambda M+t)$. Then, by construction, $M$ is exhaustive if and only if $(\lambda,t)=(1,0)$ lies on an inclusion-wise maximal set of facet-defining hyperplanes of $\Hom(M)$. The latter condition is equivalent to $(\lambda,t)=(1,0)\in\ext \Hom(M)$.
\end{proof}

\begin{proof}[Proof of \Cref{lem:exhaustive_characterization}]
	By \Cref{lem:exhaustive_no_homothetic}, $M\in \M$ is not exhaustive if and only if $M$ has a decomposition into homothets in $\mathcal{M}$, that is, there exist $M',M''\in\M$ homothetic to $M$ such that $M=\frac{1}{2}M'+\frac{1}{2}M''$.
	Suppose $\ext M=\{a\}$ is a singleton. Then, $M$ has a decomposition into homothets in $\mathcal{M}$ if and only if $a\notin\ext A$. Thus, for the remainder of the proof, assume that $\ext M$ is not a singleton. 
	
	$M$ has a decomposition into homothets in $\mathcal{M}$ if and only if one of the following holds:
	\begin{enumerate}[nosep]
		\item There exists a direction $t \in\mathbb{R}^d\setminus\{0\}$ such that $\ext M+t$ and $\ext M-t$ are both subsets of $A$ (translation).
		\item There exists a point $z\in\mathbb{R}^d$ and $\varepsilon>0$ such that $z+(1+\varepsilon) (\ext M-z)$ and $z+(1-\varepsilon) (\ext M-z)$ are both subsets of $A$ (dilation with center $z$).
	\end{enumerate}
	This is because any homothety is itself either a translation or dilation.\footnote{Specifically, suppose $M=\frac{1}{2}M'+\frac{1}{2}M''$ and $M'=z+(1+\varepsilon)(M-z)$. Plugging in and rearranging for $M''$ yields $M''=z+(1-\varepsilon)(M-z)$.} We complete the proof by showing that (1) and (2) above are equivalent, respectively, to the negations of (1) and (2) in the statement of the lemma.
	
	Suppose (1) holds. Then $t$ is orthogonal to all the normals of the hyperplanes in $\mathcal{F}(M)$ for otherwise there is a point $a\in \ext M\cap H$, for some $H\in\mathcal{F}(M)$, such that $a+t\notin H$ or $a-t\notin H$, which contradicts that $\ext M+t$ and $\ext M-t$ are subsets of $A$.  
	Hence the spanning condition $\spn \{n_H\}_{H\in\mathcal{F}(M)} =\mathbb{R}^d$ is violated. Conversely, if the spanning condition is violated, there is a direction $t\in\mathbb{R}^d\setminus\{0\}$ such that $t$ is orthogonal to all the facet normals in $\mathcal{F}(M)$. Then, $\ext M+t$ and $\ext M-t$ will still satisfy the facet-defining inequalities of $A$ for $||t||$ sufficiently small, that is, $\ext M+t,\ext M-t\subset A$.
	
	Suppose (2) holds. Consider any $H\in\mathcal{F}(M)$ and $a\in \ext M\cap H$. Then $z\in H$, for otherwise $z+(1+\varepsilon)(a-z)$ or $z+(1-\varepsilon)(a-z)$ is not in $A$. Thus, since $z\in H$ for any $H\in\mathcal{F}(M)$ such that $\ext M\cap H\neq\emptyset$, we have $\bigcap_{H\in\mathcal{F}(M)} H\neq\emptyset$. Conversely, if $\bigcap_{H\in\mathcal{F}(M)} H\neq\emptyset$, choose any $z\in \bigcap_{H\in\mathcal{F}(M)} H$. For $\varepsilon>0$ sufficiently small, $z+(1+\varepsilon) (\ext M-z)$ and $z+(1-\varepsilon) (\ext M-z)$ are both subsets of $A$. This is because $\ext M$ is uniformly bounded away from facet-defining hyperplanes $H\notin\mathcal{F}(M)$ and because $a\in H\in\mathcal{F}(M)$ implies $(z+(1\pm\varepsilon) (a-z))\in H$. 
\end{proof}

\subsection{Proofs for \texorpdfstring{\Cref{sec:main}}{Section~\ref{sec:main}}}

\label{app:main}
\subsubsection{Extreme Points for One-Dimensional Type Spaces}\label{app:extreme_1D}
In this section, we provide the complete characterization of the extreme points for one-dimensional problems promised in \Cref{sec:main}. \Cref{thm:simple} follows as a corollary. 

The key concept in the characterization, a \emph{flexible chain/loop}, requires auxiliary definitions. Fix $d=2$. For a given finite-menu mechanism $x\in\mathcal{X}$, we say that distinct $a,b\in\menu(x)$ are \textbf{adjacent} if there exists a type $\theta\in\Theta$ such that $\{a,b\}=\argmax_{\tilde a\in \menu(x)} \tilde a\cdot\theta$. Every menu item has two neighbors in the adjacency relation, except if the type space is restricted, in which case there are exactly two items $a,b\in\menu(x)$ with only one neighbor. For these two items, designate two fictitious items $*$ and $**$ and extend the adjacency relation by declaring that $*$ is adjacent only to $a$ and $**$ only to $b$.

A sequence $S=(a_1,\ldots,a_n)$ of distinct items in $\menu(x)\cup\{*,**\}$ with $n\ge 2$  is a \textbf{chain} if consecutive items are adjacent and non-consecutive items are not adjacent. It is a \textbf{loop} if $S$ lists all items in $\menu(x)$, consecutive items are adjacent, and $a_1$ and $a_n$ are also adjacent. (A loop necessitates an unrestricted type space.)

For $a\in\menu(x)\cup\{*,**\}$, we say that $a$ is \textbf{inflexible} if $a\in\ext A$; \textbf{semi-flexible} if $a\in\bndr A\setminus\ext A$ and $H\cap\menu(x)=\{a\}$, where $H$ is  the (unique) feasibility constraint of $A$ containing $a$; and \textbf{flexible} otherwise.

\begin{definition}
	A chain $S=(a_1,\ldots,a_n)$ with $n\geq 3$ is \textbf{flexible} if each $a_i$ is flexible or semi-flexible and both $a_1$ and $a_n$ are flexible. 	
    A chain $S=(a_1,a_2)$ is flexible if the line segment $\overline{a_1a_2}\not\subset\bndr A$ and both $a_1$ and $a_2$ are flexible.

	A loop $S=(a_1,\ldots,a_n)$  is \textbf{flexible} if each $a_i$ is flexible or semi-flexible and at least one of $a_1$ and $a_n$ is flexible. A loop is also flexible if each $a_i$ is semi-flexible, $n$ is even, and 
	\begin{equation}
		\label{eq:angles}
		\prod_{k=1}^n\sin\alpha_k = \prod_{k=1}^n\sin\beta_k,
	\end{equation}
	where $\alpha_k$ and $\beta_k$ are the angles depicted in the right panel of \Cref{fig:plane_extreme_points}.\footnote{Formally, fix a clock-wise ordering of the extreme points of $A$ and of the items $(a_1,\ldots,a_n)$ in $\menu(x)$. For each $k$ (indices modulo $n$), let $a_{k-1}$ and $a_{k+1}$ be the items preceding and succeeding $a_k$, respectively. Let $\overline{cd}$ be the edge of $A$ containing $a_k$, with $c$ preceding $d$. Define $\alpha_k:=\arccos\left(\frac{(a_{k-1}-a_k)\cdot(c-a_k)}{||a_{k-1}-a_k||||c-a_k||}\right)$ and $\beta_k:=\arccos\left(\frac{(a_{k+1}-a_k)\cdot(d-a_k)}{||a_{k+1}-a_k||||d-a_k||}\right)$.} 
\end{definition}

\begin{theorem}
	\label{thm:mielczarek}
	Let $d=2$ and $x\in\mathcal{X}$. Then, $x\in\ext \mathcal{X}$ if and only if either
	\begin{enumerate}
		\item $|\menu(x)|\leq 2$ and $x$ is exhaustive or
		\item $3\leq |\menu(x)|<\infty$ and $\menu(x)$ has no flexible chain or loop.
	\end{enumerate}
\end{theorem}

\begin{figure}
	\centering
	\begin{subfigure}[t]{0.45\textwidth}
		\centering
		\begin{tikzpicture}[line cap=round, line join=round, >=triangle 45, x=1.0cm, y=1.0cm]
			\coordinate (v1) at (0.,0.);
			\coordinate (v5) at (4.,1.);
			\coordinate (v4) at (5.,3.);
			\coordinate (v3) at (2.,5.);
			\coordinate (v2) at (0.,3.);
			
			\draw[thick] (0.,0.) -- (5.,0.) -- (5.,5.) -- (0.,5.) -- cycle;
			
			\draw[thick] (v1) -- (v2) -- (v3) -- (v4) -- (v5) -- cycle;
			\fill[opacity=0.04] (v1) -- (v2) -- (v3) -- (v4) -- (v5) -- cycle;

			\draw [dotted] (4.447976270795717,1.1186578317567282) -- (5.,2.1975176451092215);
			\draw [dotted] (5.,2.1975176451092215) -- (0.8052378421820068,5.);
			\draw [dotted] (0.8052378421820068,5.) -- (0.,4.190235182713238);
			\draw [dotted] (3.5468110149365755,0.890193400693848) -- (5.,3.80946113094177);
			\draw [dotted] (5.,3.80946113094177) -- (3.2168068367346367,5.);
			\draw [dotted] (3.2168068367346367,5.) -- (0.,1.8040511249453717);
			\draw [dotted] (v5) -- (4.447976270795717,1.1186578317567282);
			
			\draw [fill=black] (v1) circle (1pt);
			\draw [fill=black] (v2) circle (1pt);
			\draw [fill=black] (v3) circle (1pt);
			\draw [fill=black] (v4) circle (1pt);
			\draw [fill=black] (v5) circle (1pt);
			
			\draw[color=black] (v1) node[below left] {$a_1$};
			\draw[color=black] (v2) node[left] {$a_2$};
			\draw[color=black] (v3) node[above] {$a_3$};
			\draw[color=black] (v4) node[right] {$a_4$};
			\draw[color=black] (v5) node[below] {$a_5$};
		\end{tikzpicture}
	\end{subfigure}
	\hfill
	\begin{subfigure}[t]{0.45\textwidth}
		\centering
		\begin{tikzpicture}[line cap=round,line join=round,>=triangle 45,x=1.0cm,y=1.0cm]
			\draw [shift={(5.,2.)}] (0,0) -- (90.:1.25) arc (90.:135.:1.25) -- cycle;
			\draw [shift={(0.,3.)}] (0,0) -- (45.:1.25) arc (45.:90.:1.25) -- cycle;
			\draw [shift={(0.,3.)}] (0,0) -- (-90.:1.25) arc (-90.:-56.309932474020215:1.25) -- cycle;
			\draw [shift={(2.,5.)}] (0,0) -- (180.:1.25) arc (180.:225.:1.25) -- cycle;
			\draw [shift={(2.,5.)}] (0,0) -- (-45.:1.25) arc (-45.:0.:1.25) -- cycle;
			\draw [shift={(2.,0.)}] (0,0) -- (123.6900675259798:1.25) arc (123.6900675259798:180.:1.25) -- cycle;
			\draw [shift={(2.,0.)}] (0,0) -- (0.:1.25) arc (0.:33.690067525979785:1.25) -- cycle;
			\draw [shift={(5.,2.)}] (0,0) -- (-146.30993247402023:1.25) arc (-146.30993247402023:-90.:1.25) -- cycle;
			
			\coordinate (v1) at (2.,0.);
			\coordinate (v4) at (5.,2.);
			\coordinate (v3) at (2,5);
			\coordinate (v2) at (0,3.);
			
			\draw[thick] (0.,0.) -- (5.,0.) -- (5.,5.) -- (0.,5.) -- cycle;
			
			\draw[thick] (v1) -- (v2) -- (v3) -- (v4) -- cycle;
			\fill[opacity=0.04] (v1) -- (v2) -- (v3) -- (v4)  -- cycle;
			
			\draw [dotted] (1.6,0.)-- (5.,2.3);
			\draw [dotted] (5.,2.6)-- (2.6,5.);
			\draw [dotted] (2.6,5.)-- (0.,2.4);
			\draw [dotted] (0.,2.4)-- (1.6,0.);
			
			\draw [fill=black] (v1) circle (1pt);
			\draw [fill=black] (v2) circle (1pt);
			\draw [fill=black] (v3) circle (1pt);
			\draw [fill=black] (v4) circle (1pt);
			
			\draw[color=black] (v1) node[below] {$a_2$};
			\draw[color=black] (v2) node[left] {$a_3$};
			\draw[color=black] (v3) node[above] {$a_4$};
			\draw[color=black] (v4) node[right] {$a_1$};
			
			\begin{scriptsize}
				\draw[color=black] (1.325,0.4) node[fill=white, draw=none, inner sep=0.75pt] {$\beta_2$};
				\draw[color=black] (2.9,0.25) node[fill=white, draw=none, inner sep=0pt] {$\alpha_2$};
				
				\draw[color=black] (4.625,1.15) node[fill=white, draw=none, inner sep=0pt] {$\beta_1$};
				\draw[color=black] (4.725,2.875) node[fill=white, draw=none, inner sep=0.75pt] {$\alpha_1$};
				
				\draw[color=black] (2.925,4.675) node[fill=white, draw=none, inner sep=0pt] {$\beta_4$};
				\draw[color=black] (1.2,4.65) node[fill=white, draw=none, inner sep=0pt] {$\alpha_4$};
				
				\draw[color=black] (0.25,2.05) node[fill=white, draw=none, inner sep=0.75pt] {$\alpha_3$};
				\draw[color=black] (0.35,3.8) node[fill=white, draw=none, inner sep=0pt] {$\beta_3$};
			\end{scriptsize}
		\end{tikzpicture}
	\end{subfigure}
	\caption{An illustration of flexible chains/loops and their connection to extreme points. Left: an extended menu $M$ (shaded) with corresponding $\menu(x)=\{a_1,\ldots,a_5\}$ and flexible chain $S=(a_2,a_3,a_4,a_5)$. Dotted lines show two deformations of $M$ obtained by parallel shifts of the facet–defining lines; these deformations decompose $M$. Right: an extended menu $M$ (shaded) with $\menu(x)=\{a_1,\ldots,a_4\}$ and loop $S=(a_1,a_2,a_3,a_4)$. The loop is not flexible because it violates the symmetry condition \eqref{eq:angles} on the angles $\alpha_k$ and $\beta_k$. The dotted path depicts a candidate deformation of $M$. This candidate can be used to decompose $M$ if and only if it has no new edge orientations relative to $M$. This is not the case here ($a_1$ is split into two items) and is only possible under the symmetry condition \eqref{eq:angles}.}
	\label{fig:plane_extreme_points}
\end{figure}

\Cref{thm:mielczarek} is essentially a restatement of a result due to \citet[Theorem 3.1]{mielczarek1998extreme}. Mielczarek characterizes extremal convex bodies (i.e.,  the elements of $\ext \mathcal{M}$ when the type space is unrestricted) contained in a given convex body in the plane (i.e.,  $A$). If the type space is unrestricted ($\cone\Theta=\mathbb{R}^2$), then Mielczarek's result applies verbatim.\footnote{Specifically, in the statement of Mielczarek's result, $Q$ is our allocation space $A$, $A$ is our extended menu $M$, $A_w\cup A_p$ corresponds to the flexible menu items in $\ext M=\menu(x)$, $A_c$ corresponds to the semi-flexible menu items, and
	\begin{itemize}
		\item condition $1^\circ$ is the first condition in \Cref{thm:mielczarek}; 
		\item if $\ext M\cap\ext A\neq\emptyset$, then conditions $2^\circ$ and (i) are equivalent to the absence of a flexible chain/loop;
		\item if $\ext M\cap\ext A=\emptyset$, then conditions $2^\circ$ and (ii) are equivalent to the absence of a flexible loop.
	\end{itemize}
	Condition (iii) in Mielczarek's theorem never applies if $A$ ($Q$ in the statement) is a polytope.
}
The extension to a restricted type space, in which case the elements of $\mathcal{M}$ are closed convex sets that share the polar $\Theta^\circ$ as a common recession cone, is straightforward (see an earlier version of this paper (arXiv:2412.00649v1)).

The key idea behind the proof is that any two mechanisms that decompose a given (finite-menu) mechanism $x\in\mathcal{X}$ must make at least the same incentive and feasibility constraints binding as $x$. Geometrically, for the extended menus $M'$ and $M''$ corresponding to $x'$ and $x''$, this means that the orientations of their edges must also be edge orientations of the extended menu $M$ corresponding to $x$. Extended menus with the same edge orientations as $M$ exist if and only if $\menu(x)$ has no flexible chain or loop. \Cref{fig:plane_extreme_points} illustrates.

\begin{proof}[Proof of \Cref{thm:simple}]
	Let $M\in\ext\mathcal{M}$ be the extended menu associated with a mechanism $x\in\ext\mathcal{X}$.
	Recall that $\menu(x)=\ext M$ by \Cref{lem:isomorphisms} ($\ext M$ is closed since $d=2$); thus, we show $|\ext M|\leq |\mathcal{F}|$. 
	
	If $\cone\Theta\neq\mathbb{R}^2$, then some vertex in $\ext M$ is inflexible for otherwise $M\in\ext \mathcal{M}$ has a flexible chain. If $\cone\Theta=\mathbb{R}^2$ and no vertex in $\ext M$ is inflexible, then $M$ can only not have a flexible chain or  loop if every vertex in $\ext M$ is semi-flexible. In this case, $|\ext M|\leq |\mathcal{F}|$. Thus, we assume that $\ext M$ contains an inflexible vertex.
	
	Consider any two inflexible vertices $a,a'\in\ext M$ such that the chain $S=(a_1,\ldots,a_n)$ in between $a$ and $a'$ in the adjacency relation consists of flexible and semi-flexible allocations ($a=a'$ if there is only one inflexible vertex in $\ext M$). Without loss, assume $(a,a_1,\ldots,a_n,a')$ are ordered clockwise. Let $(H_1,\ldots,H_k)\subset\mathcal{F}$ be the sequence of feasibility constraints traversed when moving from $a$ to $a'$ clockwise along the boundary of $A$.
	
	We show that $n\leq k-1$. 
	Since $M\in\ext \mathcal{M}$, $S$ does not contain a flexible chain. Thus, $S$ contains either (1) at most one flexible vertex or (2) two adjacent flexible vertices $a_i$ and $a_{i+1}$ such that $\overline{a_ia_{i+1}}\subset H_j$ for some $j=2,\ldots,k-1$. In the first case, since $a\in H_1$ and $a'\in H_k$, there are at most $k-2$ semi-flexible vertices in $S$; hence $n\leq k-1$. In the second case, since $a\in H_1$, $a'\in H_k$, and $a_i,a_{i+1}\in H_j$, there are at most $k-3$ semi-flexible vertices in $S$; hence $n\leq k-1$.
	
	In total, we then have $|\ext M|\leq |\mathcal{F}|$. The necessity of exhaustiveness follows from \Cref{lem:exhaustive_no_homothetic}.
\end{proof}

\subsubsection{Extreme Points for Multi-Dimensional Type Spaces}\label{app:extreme_multiD}

\begin{proof}[Proof of \Cref{thm:complex}]
	The necessity of exhaustiveness follows from \Cref{lem:exhaustive_no_homothetic}. Denseness in the exhaustive mechanisms follows from \Cref{cor:dense} below. Since $\mathcal{X}$ is a compact and convex subset of a normed vector space (\Cref{lem:isomorphisms,lem:Hausdorff}), the set of extreme points is a $G_\delta$ (\citet[Lemma 7.63]{aliprantis2007infinite}). 
\end{proof}

\begin{proof}[Proof of \Cref{thm:generic_finite}]
	Immediate from \Cref{lem:indecomposable+exhaustive,lem:general_position_indecomposable}.
\end{proof}

\begin{proof}[Proof of \Cref{cor:open_dense_finite}]
	Let $x\in\mathcal{X}$ be exhaustive with associated extended menu $M\in\mathcal{M}$ such that $\menu(x)=\ext M$ is finite and non-separating. Note that general position is preserved under small perturbations of the points in $\ext M$. Openness then follows since two extended menus with the same number of extreme points are close in the Hausdorff distance if and only if the extreme points can be matched bijectively with pairwise small Euclidean distance (cf.\ \Cref{lem:Hausdorff}).
	
	We also use that non-separation is preserved under small perturbations of the points in $\ext M$. For a given extreme point $a\in\ext M$ of the polyhedron $M\in\mathcal{M}$, the normal cone 
    \begin{equation}
	\label{eq:normalcone}
	N_M(a):=\bigl\{\theta\in\cone\Theta:\ a\in\argmax_{\tilde a\in\ext M} \tilde a\cdot\theta\bigr\}
\end{equation}
 is closed. For every $\varepsilon>0$, sufficiently small perturbations of the menu items in $\ext M$ produce a polyhedron $\tilde M\in\mathcal{M}$ with corresponding extreme point $a'\in\ext\tilde M$ such that $N_{\tilde M}( a')\cap\Theta$ lies entirely within the $\varepsilon$-neighborhood of $N_M(a)\cap \Theta$. Thus, if $\Theta\setminus N_M(a)$ is connected, then $\Theta\setminus N_{\tilde M}(a')$ is connected whenever the perturbations are sufficiently small.
	
	We prove denseness by constructing extended menus $\tilde M\in\ext \mathcal{M}$ that are arbitrarily close to $M$ in the Hausdorff distance and satisfy $|\ext M|=|\ext \tilde M|$. 
	
	For this, we perturb $\ext M$ into general position while preserving exhaustiveness; call the resulting set $W$. More explicitly, if $\ext M$ is a singleton, then $M\in\ext\mathcal{M}$. If $|\ext M|\geq 2$, select an inclusion-wise minimal subset $V\subseteq\ext M$ such that the hyperplanes in $\mathcal{F}$ that contain a point of $V$ satisfy conditions (1) and (2) in  \Cref{lem:exhaustive_characterization} (this necessitates $2\le |V|\le d+1$). For each $a\in V$, if $a\in \aff(V\setminus\{a\})$, move $a$ outside $\aff(V\setminus\{a\})$ while keeping $a$ on the hyperplanes in $\mathcal{F}$ intersected by $a$, that is, in the face $F=\bigcap\{H\in\mathcal{F}\mid a\in H\}\cap A$. To see that this is possible, note that if $a$ intersects $1\leq k\leq d$ hyperplanes in $\mathcal{F}$, then $\dimension  F \geq (d-k)$ whereas $\dimension\aff(V\setminus\{a\})\leq (d-k)$ by the minimality of $V$. Also by the minimality of $V$, $F\not\subseteq \aff(V\setminus\{a\})$, hence the desired perturbation is possible. By construction, the resulting perturbed set $V$ is in general position. Now iteratively perturb the remaining points in $\ext M\setminus V$ until the resulting set $W$ is in general position. Obviously $|W|=|\ext M|$. 
	
	Define $\tilde M=\conv W+\Theta^\circ$. By construction, $\tilde M$ is a polyhedron in $\mathcal{M}$ that is arbitrarily close to $M$. For all sufficiently small perturbations, $W$ is in convex position, that is, no point of $W$ lies in the convex hull of the others, because $\ext M$ was in convex position. In addition, for all $a,a'\in W$, we have $a\notin a'+\Theta^\circ$ since the same holds for all $a,a'\in\ext M$ and since $\Theta^\circ$ is closed. Thus, $\ext\tilde M=W$. Moreover, $\tilde M$ is exhaustive by the construction of $V$. Since $\ext\tilde M$ is also in general position and non-separating, $\tilde M\in\ext\mathcal{M}$ by \Cref{thm:generic_finite}.
\end{proof}
    
\begin{proof}[Proof of \Cref{cor:dense}]
	Fix an exhaustive extended menu $M\in\M$ and $\varepsilon>0$. Choose a finite set $V\subseteq \ext M$ such that the facet-defining hyperplanes in $\mathcal{F}$ that contain a point in $V$ are exactly the hyperplanes in $\mathcal{F}(M)$ and such that, for every $a\in \ext M$, there exists $v\in V$ with $||a-v||\leq \varepsilon$. Let $M_0:=\conv V+\Theta^\circ\in\M$. By \Cref{lem:exhaustive_characterization}, $M_0$ is exhaustive. By \Cref{lem:Hausdorff}, $d(M_0,M)\leq \varepsilon$. By construction, $|\ext M_0|<\infty$.
	
	We iteratively perturb $M_0$ to eliminate all separating vertices $(a_1,\ldots,a_n)$, i.e., those vertices of $M_0$ for which $\cone\Theta\setminus N_{M_0}(a_i)$ is disconnected (cf.\ \eqref{eq:normalcone}). 
	
	Suppose $M_{i-1}\in \M$ is exhaustive, has separating vertices $(a_{i},\ldots,a_n)$, and is arbitrarily close to $M_0$. 
	
	We construct an exhaustive $M_i\in \M$ with separating vertices $(a_{i+1},\ldots,a_n)$ arbitrarily close to $M_{i-1}$. 
	Pick a hyperplane
	$$
	H_i:=\{\theta\in \cone\Theta:\; n_i\cdot \theta=0\}
	$$
	with $H_i\cap \interior N_{M_{i-1}}(a_i)\neq\emptyset$ such that both subcones
	$$
	C_i^\pm:=N_{M_{i-1}}(a_i)\cap\{\pm n_i\cdot\theta\ge 0\}
	$$
	have connected complements in $\cone\Theta$.
	If for small $\varepsilon>0$ we have $a_i+ \varepsilon n_i\in A$, set $a_{i}^{-}:=a_i$ and $a_{i}^{+}:=a_i+\varepsilon n_i$. Otherwise, if $a_i- \varepsilon n_i\in A$, set $a_{i}^{+}:=a_i$ and $a_{i}^{-}:=a_i-\varepsilon n_i$.
	If  $a_i\pm \varepsilon n_i\notin A$, then the line $L(a_i)=\{a_i+\alpha n_i:\alpha\in\R\}$ intersects $M_{i-1}$ only in the point $a_i$.
 	For $a\in A$, define the smallest face of $A$ containing $a$:
	$$
	F(a):= A \cap \bigcap_{H \in \mathcal{F}(a)}  H,
	\quad \text{where} \quad \mathcal{F}(a):= \{ H \in \mathcal{F} : a \cdot n_H = c_H \}.
	$$
	Choose $a'_i\in \interior A$ with $||a_i-a'_i||\leq\varepsilon$ such that $\dim F(a_i^-)=\dim F(a_i)+1$ or $\dim F(a_i^+)=\dim F(a_i)+1$, where $a_i^-$ and $a_i^+$ are the endpoints of the line segment $L(a_i')\cap A$. 
	Let
	$$
	M_i:=\conv\left((\ext M_{i-1}\setminus\{a_i\})\cup\{a_{i}^{-},a_{i}^{+}\}\right)+\Theta^\circ.
	$$
	By construction, $d(M_i,M_{i-1})=O(\varepsilon)$.
	Up to a set of types of measure $O(\varepsilon)$, types in $C_i^+$ select $a_{i}^{+}$, types in $C_i^-$ select $a_{i}^{-}$, and all other types make the same choices as from $M_{i-1}$. Hence $\cone\Theta\setminus N_{M_i}(a_{i}^{\pm})$ is connected by the choice of $H_i$. By construction, $$
		\spn\{n_H\}_{H\in\mathcal{F}(a_i)}=\spn \{n_H\}_{H\in\mathcal{F}(a_i^+)\cup\mathcal{F}(a_i^-)}.
		$$ 
	By \Cref{lem:exhaustive_characterization}, $M_i$ is exhaustive. Thus, $M_i$ has all the desired properties.
	
	Finally, by \Cref{cor:open_dense_finite}, $M_n$ is arbitrarily close to an element of $\ext \M$ with finite menu size, which completes the proof.
\end{proof}

\begin{proof}[Proof of \Cref{thm:infinite}]
	We begin with a few definitions. For $\tilde\M\subseteq \M$, let $\exh\tilde \M$ denote the set of exhaustive extended menus in $\tilde\M$. Let $\mathcal{B}_k\subset\exh\M$ denote the set of exhaustive extended menus that have a bounded face $f$ with $\diam(f)\geq \frac{1}{k}$ and outer unit normal vector $n_f\in\Theta$ such that $d(n_f,\bndr\cone\Theta)\geq 1/k$ (which is satisfied by convention if $\bndr\cone\Theta=\emptyset$, that is, $\cone\Theta=\mathbb{R}^d$).\footnote{The diameter of a set \( S \subseteq \mathbb{R}^d \), denoted \( \diam(S) \), is defined as
		$
		\diam(S) = \sup \{ \|a - b\| : a, b \in S \}.
		$
	}
	
	We define $\exh\M^{sc}:=\exh\M\setminus\bigcup_{k=1}^\infty \mathcal{B}_k$ and note that the mechanism  $x\in\mathcal{X}$ associated with an extended menu $M\in\exh\M^{sc}$ is continuous on
	$\interior\cone\Theta$ (in particular, $\menu(x)$ is uncountable whenever it is not a singleton). This is because, for each $M\in\exh\M^{sc}$ and $\theta\in\interior \cone \Theta$, $\argmax_{a\in M} \theta\cdot a$ is a singleton for otherwise the boundary of $M$ would contain a line segment connecting two extreme points of $M$. In particular, $x(\theta)$ is uniquely determined by $M$ on  $\interior \cone \Theta$ in the payoff-equivalence class. Therefore, the associated indirect utility function $U\in\mathcal{U}$ is differentiable on $\interior \cone \Theta$. A differentiable convex function on an open domain is continuously differentiable on that domain  (\citet[Corollary 25.5.1]{rockafellar1997convex}). Hence, $U$ is continuously differentiable on $\interior \cone \Theta$ and therefore $x=\nabla U$ is continuous on $\interior \cone \Theta$.
	
	Using the previous observation, we complete the proof by showing that $\ext\M\cap \exh\M^{sc}$ is a dense $G_\delta$ in  $\exh\M$. $\M$ is a compact metric space (\Cref{lem:Hausdorff}), hence a Baire space. The same holds for the closed, hence compact subset $\exh M\subset \M$.
	By the Baire category theorem and since $\ext \M$ is a dense $G_\delta$ in $\exh\M$ (\Cref{thm:complex}), it remains to show that $\exh\M^{sc}$ is a dense $G_\delta$ in $\exh\M$.

	By definition, $\exh\M^{sc}$ is a $G_\delta$ if each $\mathcal{B}_k$ is closed. Consider any convergent sequence $\{M_i\}_{i\in\mathbb{N}}\subset\mathcal{B}_k$ with limit $M\in\exh\M$. We show $M\in\mathcal{B}_k$. By definition, for each $i\in\mathbb{N}$, there exists a line segment $L_i\subseteq\bndr M_i$ of length $\geq \frac{1}{k}$ with normal vector $n_i\in\cone\Theta\cap\mathbb{S}^{d-1}$ such that $d(n_i,\bndr\cone\Theta)\geq \frac{1}{k}$. By selecting a subsequence if necessary, we may assume that the line segments $\{L_i\}_{i\in\mathbb{N}}$ and the normal vectors $\{n_i\}_{i\in\mathbb{N}}$
	converge to limits $L^*\subset A$ and $n^*\in\cone\Theta\cap\mathbb{S}^{d-1}$, respectively, because $\cone\Theta\cap\mathbb{S}^{d-1}$ and $A$ are compact. It is routine to verify that $L^*\subseteq \bndr M$, $L^*$ has length $\geq \frac{1}{k}$, $n^*$ is normal to $L^*$ on $\bndr M$, and $d(n^*,\bndr\cone\Theta)\geq \frac{1}{k}$. Thus, $M\in\mathcal{B}_k$.

	To show denseness of $\exh\M^{sc}$, consider the set $\exh\M\setminus\mathcal{B}_k$ for some arbitrary $k\in\mathbb{N}$. By the previous paragraph, $\exh\M\setminus \mathcal{B}_k$ is relatively open in $\exh\M$. Moreover, $\exh\M\setminus \mathcal{B}_k$ is dense in $\exh\M$ because every extended menu $M\in\exh\M$ can be approximated by an extended menu in $\exh\M$ whose bounded faces have diameter $< \frac{1}{k}$.\footnote{For example, one can mollify the support function of $M$ while preserving exhaustiveness; see \citet[Theorem 3.4.1]{schneider2014convex}.}  We have that $\exh\M^{sc}=\bigcap_{k=1}^\infty(\exh\M\setminus\mathcal{B}_k)$ is a countable intersection of relatively open and dense sets in a Baire space. Thus, by the Baire category theorem, $\exh\M^{sc}$ is dense in $\exh\M$. 
\end{proof}

\subsection{Proofs for \texorpdfstring{\Cref{sec:applications}}{Section~\ref{sec:applications}}}

\label{app:applications}
Finally, we prove the results about rationalizability for the applications to monopoly pricing, delegation and veto bargaining in \Cref{sec:applications} (\Cref{thm:monopoly,thm:delegation,thm:veto_bargaining}).

\subsubsection{Monopoly}

\begin{lemma}
	\label{lem:undomi_monopoly}
	In the monopoly problem, every IC and IR mechanism that excludes the lowest type and has no marginal distortions at the top in each dimension is undominated.
\end{lemma}

\begin{proof}
	Suppose $x$ satisfies the hypotheses but $x'$ dominates $x$. 
	
	We first show that $U'\geq U$. Extend $U$ and $U'$ to $\cone\Theta$ by 1-homogeneity. By dominance, $$\nabla_{\bar v} U'(\theta)\geq \nabla_{\bar v} U(\theta) \quad \text{for a.e. } \theta\in\cone\Theta,$$ where $\bar v =(0,\ldots,0,1)$. For a given $\theta\in\cone\Theta$ and $t\in\mathbb{R}_-$, define $z(t)=\theta+t \bar v$ and
    $h(t)=U'(z(t))-U(z(t))$.
    Note that $h$ is the difference of two convex functions restricted to the ray $z(\mathbb{R}_-)\subset\cone\Theta$, hence absolutely continuous on compact intervals and differentiable almost everywhere.
    By Fubini's theorem, for a.e. $\theta\in\cone\Theta$,  $\nabla_{\bar v} U'(z(t))\geq \nabla_{\bar v} U(z(t))$ holds for a.e. $t\in\mathbb{R}_-$. For such $t$, $\dot h(t)\geq 0$ by the chain rule, so $h$ is non-decreasing. If $\lim_{t\to-\infty} U(z(t))>0$, then $x$ would have to allocate some goods with positive probability without charging any payment, which would contradict that $x$ excludes the lowest type. Hence, $\lim_{t\to-\infty}h(t)\geq 0$. Thus, $h\geq 0$. Since this holds for a.e. $\theta\in\cone\Theta$ and indirect utility functions are continuous, $U'\geq U$, as desired.
	
	Since $x$ has no distortion at the top $\theta^*=(1,\ldots,1,-1)$, we must have $U'(\theta^*)=U(\theta^*)$. Otherwise, if $U'(\theta^*)>U(\theta^*)$, then revenue at $x'(\theta^*)$ is lower than revenue at $x(\theta^*)$ (since $x(\theta^*)$ maximizes surplus). The same would hold in an open neighborhood of $\theta^*$ by IC since our no-distortion condition requires that $\theta^*$ uniquely prefers to buy the grand bundle of all goods, contradicting dominance.

    We next construct an auxiliary objective $\tilde v:\cone\Theta\to\mathbb{R}^d$. For each $i=1,\ldots,m=d-1$, let
	$$
	\tilde v_i(\theta)=\begin{cases}
		0 &\text{if } |\theta_i/\theta_d|<1-\varepsilon_i\\
		-1+\varepsilon_i &\text{if } |\theta_i/\theta_d|\geq 1-\varepsilon_i
	\end{cases} \quad\text{and}\quad \tilde v_d(\theta)=1,
	$$
	where a small perturbation vector $\varepsilon\in\mathbb{R}^{m}_{++}$ will be needed to rigorously address non-differentiabilities.
	The objective $\tilde v$ differs from revenue-maximization only if $\theta$ is willing to pay (nearly) the maximum amount for good $i=1,\ldots,d-1$, in which case a cost of (nearly) $1$ is incurred when good $i$ is sold to $\theta$. Crucially, since $x$ has no marginal distortions at the top in each dimension (meaning type $\theta\in\Theta$ with $\theta_i=1$ strictly prefers allocations where good $i=1,\ldots,m$ is allocated with probability 1), we have 
	$$
	x'(\theta)\cdot \tilde v(\theta)\geq x(\theta)\cdot \tilde v(\theta)-\delta(\varepsilon)
	$$
    for a.e. $\theta\in\Theta$, where $\delta(\varepsilon)\to 0$ as $||\varepsilon||_\infty\to 0$
    by IC and the compactness of $\Theta$. Using a.e. differentiability and 1-homogeneity, for a.e. $\theta\in\cone\Theta$,
	\begin{equation}
		\label{eq:dominance_monopoly}
		\nabla_{\tilde v(\theta)} U'(\theta)\geq \nabla_{\tilde v(\theta)} U(\theta)-\delta(\varepsilon).
	\end{equation}

    Since $U'\neq U$ and $U'\geq U$, there exists $\theta^0\in\Theta$ such that $U'(\theta^0)>U(\theta^0)$. For the sake of clarity, we complete the proof with a slightly informal argument; the footnote below fills in the gaps. The integral curve of (a smoothed version) of the vector field $\theta\mapsto\tilde v(\theta)$ with initial condition $\theta^0$ connects $\theta^0$ to $\lambda\theta^*$ for some $\lambda\in\mathbb{R}_{++}$ as $\varepsilon\to 0$. By \eqref{eq:dominance_monopoly}, $U'(\lambda\theta^*)-U'(\lambda\theta^*)+O(\delta(\varepsilon))\geq U'(\theta^0)-U(\theta^0)>0$, a contradiction with $U'(\theta^*)=U(\theta^*)$ by the 1-homogeneity of indirect utilities.\footnote{Let us be explicit. Without loss, assume $1>\theta^0_1\geq \ldots\geq \theta^0_m>0$. For each $k=1,\ldots,m$, define
    	$$
    	\theta^{k}=\left(\frac{(1-\varepsilon_{1})\theta_k^{0}}{1-\varepsilon_{k}},\frac{(1-\varepsilon_{2})\theta_k^{0}}{1-\varepsilon_{k}},\dots,\frac{(1-\varepsilon_{k-1})\theta_k^{0}}{1-\varepsilon_{k}},\theta_k^{0},\theta_{k+1}^{0},\dots,\theta_m^{0},-\frac{\theta_k^{0}}{1-\varepsilon_{k}}\right)\in\cone\Theta.
    	$$
    	By Fubini, for a.e. $\theta^0$ and $\varepsilon\in\mathbb{R}^m_{++}$, 
    	$
    	\nabla_{\tilde v(\theta)} U'(\theta)\geq \nabla_{\tilde v(\theta)} U(\theta)-\delta(\varepsilon)
    	$
    	holds for a.e. $\theta$ along the polygonal chain $(\theta^0,\ldots,\theta^m)$. By integrating along each segment of the chain for such $\theta^0$ and $\varepsilon$, we have $U'(\theta^m)-U(\theta^m)+O(\delta(\varepsilon))\geq U'(\theta^0)-U(\theta^0)>0$ by the same arguments as in the first paragraph and the definition of $\tilde v$. Hence, by the continuity of $U$ and $U'$, as $||\varepsilon||_\infty\to 0$, we conclude $\theta_m^0(U'(\theta^*)-U(\theta^*))\geq U'(\theta^0)-U(\theta^0)>0$ since $\frac{1}{\theta_m^0}\theta^m\to\theta^*$, a contradiction.} 
\end{proof}

\begin{proof}[Proof of \Cref{lem:VtoX} (Monopoly Problem)]
Suppose not, that is, $V(\theta)=x(\theta)\cdot\bar v=x'(\theta)\cdot\bar v$ for undominated IC and IR mechanisms $x$ and $x'$. With the same arguments as in \Cref{lem:undomi_monopoly}, we get $\dot h=0$. Since $x$ and $x'$ necessarily exclude the lowest type (otherwise they allocate goods for free), $\lim_{t\to\infty} h(t)=0$. Thus, $h=0$ and $U=U'$. We conclude $x=x'$.
\end{proof}

\begin{remark}

	A mechanism that excludes the lowest type and has no distortion at the top in the usual sense can be dominated.\footnote{For example, the IC and IR mechanism that sells type $(\theta_1,\theta_2,-1)$ good 1 with probability $\tfrac34\theta_1+\tfrac14\theta_2$ and good 2 with probability $\tfrac14\theta_1+\tfrac34\theta_2$ is dominated by the IR mechanism that sells the \textit{additional} probabilities $\nabla \varepsilon(\theta_1^2+\theta_2^2)(\theta_1-\theta_2)^2$. This mechanism  is IC for all sufficiently small $\varepsilon>0$. (Payments are uniquely pinned down by revenue-equivalence and IR.)} Meanwhile, the mechanism depicted in Figure 2 of \citet{manelli2007multidimensional} is undominated, yet has distortion at the top in one dimension.  We leave a complete characterization of dominance in the multi-good monopoly problem for future work; the converse to \Cref{lem:undomi_monopoly} holds in the one-good case.

\end{remark}

\begin{proof}[Proof of \Cref{thm:monopoly}]
The first claim about weak rationalizability is immediate from \Cref{lem:undomi_monopoly,lem:undominated_optimal}. For the denseness claim, take any $M\in\mathcal{M}$ associated with an IC and IR mechanism $x\in\mathcal{X}$. Define the pricing function $p:[0,1]^m\to\mathbb{R}_+$ (cf.\ \citet[Appendix A.2]{hart2015maximal}) via
$$
p(b)=\min\{t\in\mathbb{R}_+\mid (b,t)\in M\}.
$$
Since $M$ is closed and convex, $p$ is well-defined and convex and $p(0)=0$ by IR. The form of the polar cone $\Theta^\circ$ for the monopoly problem implies $\nabla_{e_i} p(b)\in[0,1]$ for the directional derivative in unit direction $e_i$ for all $b\in[0,1)^m$ and $i=1,\ldots,m$. For all sufficiently small $\varepsilon>0$, define the pricing function $p^\varepsilon(b)=(1-2\varepsilon)p(b)+\varepsilon \mathbf{1}\cdot b$ for all $b\in[0,1]^m$ and $M^\varepsilon=\epi p^\varepsilon + \Theta^\circ$. Then, $p^\varepsilon$ is also convex, $p^\varepsilon(0)=0$, and $\nabla_{e_i} p^\varepsilon(b)=(1-2\varepsilon)\nabla_{e_i} p(b)+\varepsilon\in[\varepsilon,1-\varepsilon]$. As $\varepsilon\to 0$, the Hausdorff distance $d(\epi p, \epi p^{\varepsilon})\to 0$. By \Cref{lem:Hausdorff}, $d(M^\varepsilon,M)\to 0$. The IC and IR mechanisms $x^\varepsilon$ associated with the extended menus $M^\varepsilon$ must exclude the lowest type and have no marginal distortion at the top in each dimension because the marginal prices $\nabla_{e_i} p^\varepsilon(b)$ are bounded away from 0 and 1, so the lowest type prefers to buy nothing whereas any type with $\theta_i=1$ prefers to buy the maximum amount of good $i$.

We observe that the undominated extreme points are dense in the set of \eqref{eq:IC} and \eqref{eq:IR} mechanisms when $m\geq 2$. As in the previous paragraph, for any extended menu $M\in\mathcal{M}$ corresponding to some \eqref{eq:IC} and \eqref{eq:IR} mechanism, take an extended menu $M^\varepsilon$ with $d(M,M^\varepsilon)\leq\varepsilon$ and corresponding marginal prices in $[\varepsilon,1-\varepsilon]$, which is hence undominated. In particular, $(0,\ldots,0,0),(1,\ldots,1,p)\in\ext M^\varepsilon$ for some price $p>0$ for the grand bundle, hence $M^\varepsilon$ is exhaustive by \Cref{lem:exhaustive_characterization}. As in the proofs for \Cref{cor:open_dense_finite,cor:dense}, apply an arbitrarily small perturbation to $M^\varepsilon$ to obtain an extended menu $M^*\in \ext\M$ with $(0,\ldots,0,0),(1,\ldots,1,p)\in\ext M^*$. As long as all perturbations are sufficiently small and carried out in the price dimension, the marginal prices of $M^*$ are still bounded away from 0 and 1, hence $M^*$ is undominated.

We complete the proof by showing that the strongly rationalizable mechanisms are dense in the undominated extreme points. Fix any undominated extreme point $x\in\ext \mathcal{X}$. Let $V\in\ext\mathcal{V}$ be the associated principal's indirect utility function. \Cref{lem:v_exposed} shows that there is a sequence $(V^n)_{n\in\mathbb{N}}\subset\exp_+\mathcal{V}$ such that $V^n\to V$. By \Cref{lem:VtoX}, for each $n\in\mathbb{N}$ there is a unique IC and IR mechanism $x^n\in\mathcal{X}$ that generates $V^n$; hence each $x^n$ is strongly rationalizable (with a strictly positive belief density). Since $\mathcal{X}$ is compact, up to taking a subsequence, $x_n\to x'\in\mathcal{X}$ in $L^1$. By continuity of the map $\mathcal{X}\to\mathcal{V}$, $x'$ must generate $V$. Again by \Cref{lem:VtoX}, $x=x'$, as desired.
\end{proof}

\subsubsection{Delegation}
\begin{lemma}\label{lem:undomi_delegation}
	In the delegation problem, an IC mechanism $x$ is undominated if and only if $0\in\menu(x)$.
\end{lemma}

\begin{proof}
	If not, give the agent discretion to choose $0$ in addition to $\menu(x)$. Since $b\in\mathbb{R}_{++}$, whenever the agent prefers $0$ over $\menu(x)$, so does the principal. Since $0\in\ext A$, a positive measure of types prefers 0 to $\menu(x)$; hence the mechanism with menu $\{0\}\cup\menu(x)$ dominates $x$.

	Let $x,x'\in\mathcal{X}$  with indirect utilities $U,U'\in\mathcal{U}$ and assume $0\in \menu(x)$. Suppose for contradiction that $x'$ dominates $x$, that is,
	\[
	(\theta-b)\cdot\bigl(x'(\theta)-x(\theta)\bigr)\ \ge\ 0
	\quad\text{for a.e.\ }\theta\in\Sph,
	\]
	with strict inequality on a set of positive measure. By the same argument as in the first paragraph, $0\in \menu(x')$. Since $x\in\partial U$ and $x'\in\partial U'$ and using 1-homogeneity of indirect utilities,
	\begin{equation}\label{eq:dom}
		\frac{U'(\theta)-U(\theta)}{||\theta||}\ \ge\ \nabla_b\bigl(U'(\theta)-U(\theta)\bigr)
		\qquad\text{for a.e.\ }\theta\in\mathbb{R}^d.
	\end{equation}
	
	Fix a generic $\theta\in\Sph$ and consider the ray $z(t):=\theta+tb$ with $t\in\R$ and its norm $r(t):=\|z(t)\|$ (which is bounded away from 0 for $\theta$ not collinear with $b$). Define
	\[
	h(t)\ :=\ U'\bigl(z(t)\bigr)-U\bigl(z(t)\bigr).
	\]
	Note that $h$ is the difference of two convex functions restricted to a line, hence absolutely continuous on compact sets and differentiable almost everywhere.
	For a.e. $\theta\in\Sph$, \eqref{eq:dom} holds a.e. along the ray $z(t)$ (Fubini) and, by the chain rule, 
	\begin{equation}\label{eq:ode}
		\dot h(t)\ \le\ \frac{h(t)}{r(t)} \quad\text{for a.e. } t\in\R.
	\end{equation}
	
	Because $0\in \menu(x)\cap \menu(x')$, there exists $T\in\mathbb{R}$ such that $U(z(t))=U'(z(t))=0$ for all $t\leq T$.
	Let
	\[
	\Psi(t) := \exp\left(-\int_{T}^{t}\frac{ds}{r(s)}\right)\,h(t),
	\]
	so that
	\[
	\dot\Psi(t) = \exp\left(-\int_{T}^{t}\frac{ds}{r(s)}\right)\left(\dot h(t)-\frac{h(t)}{r(t)}\right)\ \le\ 0
	\quad\text{for a.e.\ }t.
	\]
	Hence $\Psi(t)$ is nonincreasing. By construction, $\Psi(T)=0$. Thus, $\Psi\leq 0$ and $h\leq 0$, so $h$ is also nonincreasing by \eqref{eq:ode}. Since $\theta\in\mathbb{S}^{d-1}$ was generic and $U$ and $U'$ are continuous, we have $U'\leq U$.
	
		As an intermediate step, we note $\lim_{t\to\infty}\frac{h(t)}{r(t)}=0$ whenever $||b||<1$. 
	Since $U'\leq U$ and $x'$ dominates $x$,
	we have
	$$
	1\leq \frac{x'(\theta^*)\cdot(\theta^*-b)}{x(\theta^*)\cdot(\theta^*-b)}=\frac{U'(\theta^*)}{U(\theta^*)}\leq 1,
	$$
	so $U'(\theta^*)=U(\theta^*)$. Continuity and 1-homogeneity of $U$ and $U'$ imply 
	$$
	\lim_{t\to\infty}\frac{h(t)}{r(t)}=\lim_{t\to\infty} (U'-U)\left(\frac{\theta+tb}{||\theta+tb||}\right)=0.
	$$
	
	To conclude, we show $h\equiv 0$. Since $h\leq 0$ and $h$ is non-increasing, if $h\not\equiv 0$, then $h(t)<0$ for all sufficiently large $t$. Note that $\dot r(t)=\frac{(\theta+tb)\cdot b}{||\theta+tb||}\to||b||\in(0,1)$ as $t\to\infty$. Thus, for all sufficiently large t,
	$$
	\frac{d}{dt}\left(\frac{h}{r}\right)(t)=\frac{\dot h r-h\dot r}{r^2}\leq \frac{\frac{h}{r}r-h\dot r}{r^2}=\frac{(1-\dot r)h}{r^2}<0.
	$$
	Since $h/r<0$, we contradict  $\lim_{t\to\infty}\frac{h(t)}{r(t)}=0$. Thus, $h\equiv 0$. Since $\theta\in\mathbb{S}^{d-1}$ was generic and $U$ and $U'$ are continuous, we have $U'= U$, so $x'=x$, a contradiction.
\end{proof}

\begin{proof}[Proof of \Cref{lem:VtoX} (Delegation Problem)]
    Suppose not, that is, $V(\theta)=x(\theta)\cdot v(\theta)=x'(\theta)\cdot v(\theta)$ for undominated IC mechanisms $x$ and $x'$. Then, in the proof of \Cref{lem:undomi_delegation}, \eqref{eq:dom} and \eqref{eq:ode} hold with equality. Hence $\Psi=0$ implies $h=0$ implies $U=U'$. We conclude $x=x'$.
\end{proof}

The proof for \Cref{thm:delegation} is analogous to the proof of \Cref{thm:monopoly} and hence omitted.

\subsubsection{Veto Bargaining}

\begin{lemma}\label{lem:undomi_veto}
	In the veto bargaining problem, an IC and IR mechanism $x$ is undominated if and only if $\ubar a,a^*\in\menu(x)$.
\end{lemma}

\begin{proof}
	We first show that the conditions given in the statement are necessary. For this, fix any IC and IR mechanism $x$. It is clear that $\ubar a\in\menu(x)$ for otherwise $x$ is not IR since there is a type $\ubar\theta$ for which $\ubar a$ is their (unique) most preferred alternative in $A$. If $\menu(x)$ does not contain the principal's favorite alternative $a^*\in\ext A$, obtain a new IC and IR mechanism $x'$ by letting the agent choose from $\menu(x)\cup\{a^*\}$. Thus, for all $\theta\in\Theta$, either $x'(\theta)=x(\theta)$ or $x'(\theta)=a^*$. Since $a^*\in\ext A$, there is a positive measure of types for which $a^*$ is their most preferred allocation in $A$. Thus, $x'$ dominates $x$.
	
	For sufficiency, suppose $x$ is an IC and IR mechanism such that $\ubar a, a^*\in \menu(x)$. For the sake of contradiction, suppose the IC and IR mechanism $x'$ dominates $x$. By the same argument as in the previous paragraph, $\ubar a,a^*\in\menu(x')$. Let $U$ and $U'$ be the agent's indirect utility functions associated with $x$ and $x'$, respectively. 
	
	We complete the proof. Since $x\in\partial U$ and $x'\in\partial U'$ and using 1-homogeneity of indirect utilities and dominance,
\begin{equation}\label{eq:dom2}
		 \nabla_b\bigl(U'(\theta)-U(\theta)\bigr)\geq 0
		\qquad\text{for a.e.\ }\theta\in\mathbb{R}^d.
	\end{equation}
	Fix a generic $\theta\in\Sph$ and consider the ray $z(t):=\theta+tb$ with $t\in\R$. Define
	$$
	h(t)\ :=\ U'\bigl(z(t)\bigr)-U\bigl(z(t)\bigr).
	$$
	Note that $h$ is the difference of two convex functions restricted to a line, hence absolutely continuous on compact sets and differentiable almost everywhere.
	For a.e. $\theta\in\Sph$, \eqref{eq:dom2} holds a.e. along the ray $z(t)$ (Fubini) and, by the chain rule, 
	$\dot h(t)\ \ge 0$ for a.e. $t\in\R$.
	Because $\ubar a,a^*\in \menu(x)\cap \menu(x')$, there exists $T\in\mathbb{R}_+$ such that $U(z(t))=U'(z(t))=0$ for all $t\leq -T$ and $U(z(t))=U'(z(t))=a^*\cdot z(t)$ for all $t\geq T$. Hence, $h(t)=0$ for all $t\leq -T$ and $t\geq T$. Since $\dot h(t)\ \ge 0$, $h\equiv 0$. Since $\theta\in\mathbb{S}^{d-1}$ was generic and $U$ and $U'$ are continuous, we have $U'= U$, so $x'=x$, a contradiction.
\end{proof}

\begin{proof}[Proof of \Cref{lem:VtoX} (Veto Bargaining Problem)]
    Immediate from the proof of \Cref{lem:undomi_veto}.
\end{proof}

The proof for \Cref{thm:veto_bargaining} is analogous to the proof of \Cref{thm:monopoly} and hence omitted.

\subsection{Proofs for \texorpdfstring{\Cref{sec:proof_sketches}}{Section~\ref{sec:proof_sketches}}}
\label{app:technical}

\subsubsection{Mechanism-Menu-Utility Equivalence}
\label{app:isomorphisms}

We need the following lemma.

    \begin{lemma}
		\label{lem:Hausdorff}
		$\mathcal{M}$ equipped with the Hausdorff distance $d$ 
		is a compact metric space and $d(M,M')\leq d(\conv\ext M,\conv \ext M')$. 
	\end{lemma}
	\begin{proof}[Proof of \Cref{lem:Hausdorff}]
		
        We have
		\begin{align*}
			d(M,M')&=\inf\left\{\varepsilon>0:\:
			M\subseteq M'+\varepsilon B  \text{ and } 
			M'\subseteq M+\varepsilon B
			\right\}\\
			&=\inf\left\{\varepsilon>0:\:
			\begin{multlined}
				\conv\ext M+\Theta^\circ\subseteq \conv\ext M'+\Theta^\circ+\varepsilon B  \text{ and } \\
				\conv\ext M'+\Theta^\circ\subseteq \conv\ext M+\Theta^\circ+\varepsilon B
			\end{multlined}
			\right\}\\
			&\leq \inf\left\{\varepsilon>0:\:
			\begin{multlined}
				\conv\ext M\subseteq \conv\ext M'+\varepsilon B  \text{ and } \\
				\conv\ext M'\subseteq \conv\ext M+\varepsilon B
			\end{multlined}
			\right\}\\
			&=d(\conv\ext M,\conv \ext M')<\infty,
		\end{align*}
		where 
		the first inequality is because  $Z_1\subseteq Z_2$ implies $Z_1+Z_3\subseteq Z_2+Z_3$ for $Z_1,Z_2,Z_3\subset\mathbb{R}^d$, and the second inequality is because $A$ is compact. 
		
		We conclude that $(\mathcal{M},d)$ is a metric space since extended menus are closed, since the Hausdorff distance is an extended metric on the space of all closed subsets of $\mathbb{R}^d$, and since the distance is finite by the first paragraph. 
	
	It remains to show that $(\mathcal{M},d)$ is compact. Consider any sequence $\{M_n\}_{n\in\mathbb{N}}\subset\M$. By Blaschke's selection theorem, 
	$\{\cconv\ext M_n\}_{n\in\mathbb{N}}$ has a convergent subsequence $\{\cconv\ext M_{n_k}\}_{k\in\mathbb{N}}$ with compact convex limit $K\subseteq A$.  Let $M=K+\Theta^\circ\in\M$. 
	By the first paragraph, the subsequence $\{ M_{n_k}\}_{k\in\mathbb{N}}$  convergences to $M\in\mathcal{M}$. Thus, $\M$ is compact.  
	\end{proof}

\begin{proof}[Proof of \Cref{lem:isomorphisms}]
	We first show that $\Phi_1,\Phi_2,$ and $\Phi_3$ are well defined and bijective.
	
	The map $\Phi_1:\mathcal{X}\to\mathcal{M}$ is well-defined. To see this, let $M=\conv\menu(x)+\Theta^\circ$. 
	We have
	$
	\ext M=\ext(\conv \menu(x)+\Theta^\circ)\subseteq \ext(\conv \menu(x))\subset A
	$
	since $\Theta^\circ$ is a cone. $M$ is closed and convex because it is a sum of a compact convex set and a closed convex set.
	
	The map $\Phi_2:\mathcal{M}\to\mathcal{U}$ is also well-defined. To see this, observe that the support function 
	$U:z\mapsto\sup_{y\in M} y\cdot z$ of $M\in\mathcal{M}$ is an indirect utility function in $\mathcal{U}$. This is because $\sup_{y\in M} y\cdot z <\infty$ if and only if $z\in \cone\Theta$ since $\recc M = \Theta^\circ$. In this case, $\argmax_{a\in \ext M} a\cdot\theta\subset A$ is non-empty for all $\theta\in\Theta$ and every selection from the argmax defines an IC mechanism with indirect utility function $U$. 

    We next show that the composition $\Phi_2\circ\Phi_1$ assigns to each IC mechanism $x\in\mathcal{X}$ its indirect utility function $U\in\mathcal{U}$. This is because
	$$
	U(\theta)
	=
	\max_{a\in \menu(x)} a\cdot \theta 
	= \max_{y\in \conv \menu(x)+\Theta^\circ} y\cdot \theta
	= ((\Phi_2\circ\Phi_1)(x))(\theta).
	$$ 
    To see the first equality, note that by the definition of the essential range, $x(\theta)\in\argmax_{a\in \menu(x)} a\cdot \theta $ for a.e. $\theta\in\Theta$. Thus, $U(\theta)
	=\max_{a\in \menu(x)} a\cdot \theta$ for a.e. $\theta\in\Theta$. Since $U$ is continuous and the essential range $\menu(x)$ is closed (support of a measure on $A$), the equality holds for every $\theta\in\Theta$.

    We observe that the composition $\Phi_2\circ\Phi_1$ is bijective. It is injective by the definition of payoff-equivalence and surjective since $\mathcal{U}$ is defined as the set of all indirect utility functions associated with the mechanisms in $\mathcal{X}$.

	We can now conclude that
	$\Phi_1:\mathcal{X}\to\mathcal{M}$ and $\Phi_2:\mathcal{M}\to\mathcal{U}$ are bijective because $\Phi_2\circ\Phi_1$ is bijective and $\Phi_2$ is injective. The latter holds since support functions $U$ defined on $\mathbb{R}^d$, where we set $U(z)=\infty$ for $z\notin \cone\Theta$, uniquely determine their associated closed convex sets (\citet[Theorem V.2.2.2]{hiriart1996convex}). 
    
	It remains to verify that the inverse of $\Phi_2\circ\Phi_1:\mathcal{X}\to\mathcal{U}$ is given by  $\Phi_3:\mathcal{U}\to\mathcal{X}$. Take any mechanism $x\in\mathcal{X}$ with $M=\Phi_1(x)\in\mathcal{M}$ and $U\in\Phi_2(M)\in\mathcal{U}$. We have $x(\theta)\in \argmax_{a \in M} a \cdot \theta$ since the indirect utility function $U$ is the support function of $M$. $\partial U(\theta)=\argmax_{a \in M} a \cdot \theta$ is a standard property of support functions (\citet[VI, Equation (3.1)]{hiriart1996convex}). 
	
	We next show that $\Phi_1,\Phi_2,$ and $\Phi_3$ are affine. That $\Phi_2:\M\to\mathcal{U}$ commutes with convex combinations is standard property of support functions (\citet[Theorem V.3.3.3]{hiriart1996convex}). That $\Phi_3:\mathcal{U}\to\mathcal{X}$ commutes with convex combinations follows immediately from the linearity of the gradient map, which is defined almost everywhere. Thus, $\Phi_1:\mathcal{X}\to\mathcal{M}$ must also commute with convex combinations.

	A sequence of closed convex sets converges in the Hausdorff distance if and only if the associated support functions converge with respect to the sup norm on any compact set (\citet[Theorem 6]{salinetti1979convergence}). Thus, since the Hausdorff distance between any two extended menus in $\mathcal{M}$ is finite (\Cref{lem:Hausdorff}), $\Phi_2$ is a homeomorphism. Since $\mathcal{M}$ is compact, $\mathcal{U}$ is also compact.
	
	We next show that $\Phi_3$ is continuous. Consider any convergent sequence $\{U_n\}_{n\in\mathbb{N}}\subset\mathcal{U}$ with limit $U\in\mathcal{U}$. Let $D_n\subset\cone\Theta$ be the set of points where $U_n$ is differentiable, and let $D\subset\cone\Theta$ be the set of points where $U$ is differentiable (extend indirect utility functions to $\cone\Theta$ using 1-homogeneity). Let $D^*=D\cap\bigcap_{n\in\mathbb{N}} D_n$. Indirect utility functions are convex and therefore almost everywhere differentiable. Moreover, the countable union of nullsets is null; thus $\cone\Theta\setminus D^*$ is null. We have $\nabla U_n(\theta) \to \nabla U(\theta)$ for all $\theta\in D^*$ (\citet[Theorem VI.6.2.7]{hiriart1996convex}). Thus, $x_n=\Phi_3(U_n)\to \Phi_3(U) = x$ pointwise almost everywhere. Since support functions are sublinear, the Dominated Convergence theorem implies convergence in $L^1$.
	
	Finally, a continuous bijection $\Phi_3:\mathcal{U}\to\mathcal{X}$ from a compact space to a Hausdorff space is a homeomorphism; consequently, $\Phi_1=(\Phi_3\circ\Phi_2)^{-1}$ is a homeomorphism.

It remains to show that for an IC mechanism $x\in\mathcal{X}$ with associated extended menu $M=\Phi_1(x)\in\M$ and associated indirect utility function $U=\Phi_2(M)\in\mathcal{U}$, $$\menu(x)=\cl\{\nabla U(\theta):\: U \text{ is differentiable at } \theta\}=\cl\ext M.$$

        For the first equality, let $D=\{\theta\in\Theta:\: U \text{ is differentiable at } \theta\}$. 
        Suppose $a\notin\cl\nabla U(D)$. Then, $\nabla U^{-1}(\tilde A)=\emptyset$ for any sufficiently small open neighborhood $\tilde A$ of $a$. Since $x=\nabla U$ almost everywhere, $x^{-1}(\tilde A)$ is a nullset. Hence, $a\notin \menu(x)$. 
	       Conversely, suppose $a\in\cl\nabla U(D)$. Let $\tilde A$ be any open neighborhood of $a$. Then there exists $\theta^*\in D$ with $\nabla U(\theta^*)\in\tilde A$. Choose an open set $B$ such that $\nabla U(\theta^*)\in B\subset\tilde A$. Since $\partial U$ is upper hemi-continuous (\citet[Theorem VI.6.2.4]{hiriart1996convex}) and $\partial U(\theta^*)=\{\nabla U(\theta^*)\}$, there exists an open neighborhood $\tilde\Theta$ of $\theta^*$ such that $\partial U(\theta)\subset B$ for all $\theta\in\tilde\Theta$. In particular, for almost every $\theta\in\tilde\Theta$, $\nabla U(\theta)\in B\subset\tilde A$. Hence $x^{-1}(\tilde A)$ has positive measure. Since $\tilde A$ was arbitrary, $a\in\menu(x)$.
    
        For the second equality, $\partial U(\theta)=\{a^*\}$, that is, $U$ is differentiable at $\theta$, if and only if $\max_{a\in M} a\cdot\theta$ has the unique solution $a^*$ (\citet[VI, Equation (3.1)]{hiriart1996convex}). Thus, $a^*$ is an exposed point of $M$. Conversely, any exposed point $a^*$ of $M$ is the unique maximizer of $\max_{a\in M} a\cdot\theta$ for some $\theta\in\Theta$. Hence, $\{\nabla U(\theta):\: U \text{ is differentiable at } \theta\}=\exp M$. Straszewicz's theorem shows that $\ext M\subseteq\cl\exp M$ for a closed convex set $M$.
    \end{proof}

\subsubsection{Indecomposability}
\label{app:indecomposability}

\begin{proof}[Proof of \Cref{lem:indecomposable+exhaustive}]
	We show the converse assuming that $A$ is the unit simplex and $\Theta$ is unrestricted. In this case, extended menus are exactly the convex bodies (non-empty compact convex sets) contained in the allocation space. By \Cref{lem:exhaustive_no_homothetic}, if $M$ is not exhaustive, then $M\notin\ext\mathcal{M}$. Thus, it remains to show that if $M$ is decomposable, then $M\notin\ext\mathcal{M}$. 

	For any convex body  $K\subset\mathbb{R}^d$, let
	$$
	\Delta(K)=\left\{z\in\mathbb{R}^d\ \middle \vert \ \sum_{i=1}^d z_i\leq \max_{a\in K} \sum_{i=1}^d a_i    \right\}\cap\bigcap_{i=1}^d \left\{z\in\mathbb{R}^d\ \middle \vert \  z_i\geq \min_{a\in K} a_i    \right\}
	$$
	denote the smallest simplex that contains $K$ and is homothetic to the allocation simplex $A$. $\Delta(\cdot)$ commutes with Minkowski addition because the function $K\mapsto \max_{z\in K} z\cdot\theta$ commutes with Minkowski addition for any $\theta\in\mathbb{R}^d$ and because the non-empty intersection of parallel translates of facet-defining halfspaces of a simplex is homothetic to that simplex. 
	
	Suppose the extended menu $M \in \mathcal{M}$ is decomposable; that is, there exist convex bodies $K', K'' \subset \mathbb{R}^d$ not homothetic to $M$ such that $M = K' + K''$. By the previous paragraph,
	\begin{equation}
		\label{eq:simplices}
		\Delta(M)=\Delta(K')+\Delta(K'')\subseteq A.
	\end{equation}
	By definition of $\Delta(\cdot)$, there exist $s,t\in\mathbb{R}^d$ and $\lambda,\rho>0$ such that 
	\begin{equation}
		\label{eq:simplex_transform}
		\dfrac{\Delta(K')-s}{\lambda}=\Delta(M)=\dfrac{\Delta(K'')-t}{\rho}
	\end{equation}
	Solving \eqref{eq:simplex_transform} for $\Delta(K')$ and $\Delta(K'')$, respectively, and plugging into \eqref{eq:simplices} yields $s+t=0$ and $\rho+\lambda=1$.
	Now,
	$$
	M=K'+K''=\lambda\underbrace{\dfrac{K'-s}{\lambda}}_{=M'} + (1-\lambda)\underbrace{\dfrac{K''+s}{1-\lambda}}_{=M''}. 
	$$
	$M',M''\in\mathcal{M}$ are convex bodies in $A$ not homothetic to $M$ because $\Delta(M')=\Delta(M'')=\Delta(M)\subseteq A$.
\end{proof}

\begin{proof}[Proof of \Cref{lem:plane_indecomposable}]
	For the unrestricted type space, that is, $\cone\Theta=\mathbb{R}^2$, extended menus are convex bodies. \citet{meyer1972decomposing} and \citet{silverman1973decomposition} show that the indecomposable convex bodies in the plane are exactly points, line segments, and triangles; hence $|\ext M|\leq 3$. For a restricted type space, the same bound $|\ext M|\leq 3$ holds trivially because a mechanism/extended menu that is not an extreme point on the unrestricted type space is not an extreme point on any restricted type space either. Suppose, then, the type space is restricted. Since $|\ext M|\leq 3$, $\ext M$ is in general position. Moreover, $\ext M$ is non-separating if and only if $|\ext M|\leq 2$. \Cref{lem:general_position_indecomposable} completes the proof. 
\end{proof}

\begin{proof}[Proof of \Cref{lem:general_position_indecomposable}]
	Suppose $M\in\mathcal{M}$ and $\ext M$ with $|\ext M|<\infty$ is separating the type space. Pick $a\in\ext M$ so that $\{\theta\in\cone\Theta:\ a\notin\argmax_{\tilde a\in \ext M}\tilde a\cdot\theta\}$ is disconnected with components $\{\theta\in\cone\Theta:\ \argmax_{\tilde a\in \ext M}\tilde a\cdot\theta\cap(A^2\cup\{a\})=\emptyset\}$ and $\{\theta\in\cone\Theta:\ \argmax_{\tilde a\in \ext M}\tilde a\cdot\theta\cap(A^1\cup\{a\})=\emptyset\}$, where $A^1\cup A^2\cup\{a\}=\ext M$. For $\varepsilon>0$ sufficiently small, set $$A^1_\pm:=\{a+(1\pm\varepsilon)(b-a):\ b\in A^1\} \quad \text{and} \quad M_\pm:=\conv(A^1_\pm\cup A^2\cup\{a\})+\Theta^\circ\in\mathcal{K}.$$ Define $T_\pm:\ext M\to\ext M_\pm$ by $T_\pm(a)=a$, $T_\pm(b)=a+(1\pm\varepsilon)(b-a)$ for $b\in A^1$, and $T_\pm(c)=c$ for $c\in A^2$.

We claim that, for every $\theta\in\Theta$, the maximizers in $M_\pm$ are the image under $T_\pm$ of the maximizers in $M$, that is,
\[
\argmax_{y\in\ext M_\pm}y\cdot\theta = T_\pm\big(\argmax_{y\in\ext M}y\cdot\theta\big).
\]
This is because: (i) within $A^1$, $\argmax_{b\in A^1}(a+(1\pm\varepsilon)(b-a))\cdot\theta=\argmax_{b\in A^1}(b-a)\cdot\theta$; (ii) points in $A^2\cup\{a\}$ are fixed under $T_\pm$; (iii) for any $b\in A^1$, $(a+(1\pm\varepsilon)(b-a))\cdot\theta-a\cdot\theta=(1\pm\varepsilon)(b-a)\cdot\theta$, so the sign of $b\cdot\theta-a\cdot\theta$ is preserved; (iv) by the choice of $a$, there are no $A^1$–$A^2$ ties.
Using the isomorphism between extended menus and mechanisms/choice functions (\Cref{lem:isomorphisms}), we conclude $M=\frac{1}{2}M_+ +\frac{1}{2}M_-$. Since $M_\pm$ are not homothetic to $M$, $M$ is not indecomposable.

For the converse, suppose $M\in\mathcal{M}$ and $\ext M$ with $|\ext M|<\infty$ is in general position and non-separating. If $\cone\Theta=\mathbb{R}^d$, then $M$ is a polytope and, by general position of $\ext M$, it is a simplicial polytope. We can therefore apply \citet[Theorem 13]{shephard1963decomposable} to conclude that $M$ is indecomposable. Thus, henceforth suppose $\cone\Theta\neq\mathbb{R}^d$, which implies $0\notin\interior\cone\Theta$.

Say that $a,b\in\ext M$ are adjacent if $\{a,b\}=\argmax_{\tilde a\in \ext M} \tilde a\cdot\theta$ for some $\theta\in\interior\cone\Theta$; equivalently, if the normal cones $N_M(a)$ and $N_M(b)$ intersect in a common $(d-1)$-dimensional face, where we remind the reader that $N_M(a)=\bigl\{\theta\in\cone\Theta:\ a\in\argmax_{\tilde a\in\ext M} \tilde a\cdot\theta\bigr\}.$

We observe that $\interior\cone\Theta\setminus N_M(a)$ is path-connected.
Since $\ext M$ is non-separating, $\cone\Theta\setminus N_M(a)$ is connected; in fact, path-connected because both sets are closed and convex. Any path connecting $\theta,\theta'\in\interior\cone\Theta$ through $\cone\Theta$ admits a uniform neighborhood that does not intersect $N_M(a)$ because a path is compact and $N_M(a)$ is closed. Hence, $\interior\cone\Theta\setminus N_M(a)$ is path-connected.

We next show that $\bndr N_M(a)\cap\interior\cone\Theta$ is path-connected. Fix any arbitrary $\theta^*\in\interior N_M(a)$, which is non-empty because normal cones to extreme points are full-dimensional. Define the gauge
$$
\phi(\theta)=\inf\{\lambda>0:\: (\theta-\theta^*)\in \lambda(N_M(a)-\theta^*)\}.
$$
Since the gauge of the closed convex set $(N_M(a)-\theta^*)$ is continuous (e.g.,  \citet[Theorem V.1.2.5]{hiriart1996convex}), the map $\interior\cone\Theta\setminus N_M(a)\to\bndr N_M(a)\cap \interior\cone\Theta$ given by
$
\theta\mapsto\theta^*+\frac{\theta-\theta^*}{\phi(\theta)}
$
is continuous. It has values in $\bndr N_M(a)$ by construction and in $\interior\cone\Theta$ because $\interior\cone\Theta$ is convex. Fix any $\theta,\theta'\in\bndr N_M(a)\cap\interior\cone\Theta$. Let $\varepsilon>0$ be so small that $\theta+(1+\varepsilon)(\theta-\theta^*),\theta'+(1+\varepsilon)(\theta'-\theta^*)\in\interior\cone\Theta$. By the previous paragraph, find a path $\gamma:[0,1]\to \interior\cone\Theta\setminus N_M(a)$ such that $\gamma(0)=\theta+(1+\varepsilon)(\theta-\theta^*)$ and $\gamma(1)=\theta'+(1+\varepsilon)(\theta'-\theta^*)$. Then the image of $\gamma$ under the continuous map $
\theta\mapsto\theta^*+\frac{\theta-\theta^*}{\phi(\theta)}
$ is a path in $\bndr N_M(a)\cap\interior\cone\Theta$ from $\theta$ to $\theta'$.

We further observe that any two points in $\bndr N_M(a)\cap\interior\cone\Theta$ can be connected by a path that does not traverse any face of $N_M(a)$ of dimension $d-3$ or lower (except at the endpoints) because $\bndr N_M(a)\cap\interior\cone\Theta$ is relatively open in $\bndr N_M(a)$.

Let $\mathcal{T}$ denote the set of bounded $2$-dimensional faces of $M$, that is, triangular faces of $M$ since $\ext M$ is in general position. We say that $\mathcal{T}'\subseteq\mathcal{T}$ is strongly connected if for any two triangles $\Delta_1,\Delta_n\in\mathcal{T}'$  there exists a sequence of triangles $(\Delta^1,\Delta^2,\ldots,\Delta^{n-1},\Delta^n)$, each in $\mathcal{T}'$, such that consecutive triangles share a common edge.

We next claim that if $\ext|M|\geq 3$, then every $a\in \ext M$ has an adjacent vertex $b\in\ext M$ such that $\mathcal{T}(a)\cap\mathcal{T}(b)\neq\emptyset$. This is because $a$ has a least two neighbors, for otherwise the removal of $N_M(b)$ disconnects $\cone \Theta$, so the path connecting the corresponding $(n-1)$-dimensional faces of $N_M(a)$ can be chosen to traverse a common $(n-2)$-dimensional face of $a$ and $b$, which corresponds to a triangle containing both $a$ and $b$.

Let $\mathcal{T}(a)\subseteq\mathcal{T}$ denote the subset of triangles containing the vertex $a\in \ext M$; we show that $\mathcal{T}(a)$ is strongly connected. Triangles containing $a$ correspond to $(d-2)$-dimensional faces of $N_M(a)$ contained in $\interior\cone\Theta$, and two triangles share an edge containing $a$ if the corresponding $(d-2)$-dimensional faces of $N_M(a)$ are contained in a common $(d-1)$-dimensional face of $N_M(a)$. Consider any path joining two $(d-2)$-dimensional faces in $\bndr N_M(a)\cap\interior\cone\Theta$ that traverses no faces of dimension $(d-3)$ or less. The $(d-2)$-dimensional faces traversed along this path correspond to the desired sequence of triangles.

Suppose $\mathcal{T}\neq\emptyset$. By the previous two paragraphs, $\mathcal{T}(a)$ is non-empty for all $a\in \ext M$ and $\mathcal{T}$ is strongly connected. \citet[Theorem 5.1]{smilansky1987decomposability}  implies that $M$ is indecomposable.\footnote{Smilansky shows that a polyhedron is indecomposable if it has a strongly connected set of triangular faces that touches all the facets of the polyhedron. Here, ``touches'' means ``has non-empty intersection with;'' this is clear from the duality relationship described in the first paragraph of the proof.}
Lastly suppose $\mathcal{T}=\emptyset$. Then necessarily $|\ext M|\le 2$, in which case $M$ is indecomposable.
\end{proof}

\subsubsection{Dominance and Rationalizability: Preliminaries}

We prove \Cref{lem:v_exposed} and obtain \Cref{lem:undominated_optimal} as a corollary along the way. 

We define the following subsets of $\mathcal{V}$: $\und\mathcal{V}\subset \mathcal{V}$ is the set of undominated principal utility functions;
 $\underline{\und}\mathcal{V} \subset \und\mathcal{V}$ is the set of undominated principal utility functions that are strictly suboptimal for every probability density $f\in L^\infty(\Theta)$ that is uniformly bounded away from 0;
$\exp_+ \mathcal{V}\subset \mathcal{V}$ is the set of principal utility functions that are \emph{uniquely} optimal for some probability density $f\in L^\infty(\Theta)$ that is uniformly bounded away from 0. Note $\exp_+ \mathcal{V}\subset \und\mathcal{V}\cap\ext\mathcal{V}$. 
We write $\inner{V}{f}=\int_\Theta V(\theta)f(\theta)\,d\theta$.

We prove \Cref{lem:v_exposed} in three steps. The argument for the first step is inspired by (but not implied by) the argument for Theorem 9 in \citet{manelli2007multidimensional}; note the correction in \citet{manelli2012multidimensional}.

\begin{lemma}
	\label{lem:undominated1}
	$\underline{\und}\mathcal{V}\subseteq \cconv (\exp_+\mathcal{V})$.
\end{lemma}

\begin{proof}
	Fix any $V\in \underline{\und}\mathcal{V}$. We show the claim by constructing a convergent sequence of points in $\mathcal{V}$ that are convex combinations of points in $\exp_+\mathcal{V}$ with limit $V$.
	
	For $\varepsilon\geq 0$, let
	$$
	F_\varepsilon=\{f\in L^\infty(\Theta)\mid \varepsilon \leq f \leq 1\}.
	$$
	Up to renormalization, these functions are essentially bounded probability densities that are uniformly bounded away from zero. By the Banach-Alaoglu theorem, $F_\varepsilon$ is weak*-compact because it is a weak*-closed subset of the dual unit ball. 
	
	Recall that $V\in \underline{\und}\mathcal{V}$, that is, $V$ is strictly suboptimal for every density $f\in L^\infty(\Theta)$ that is uniformly bounded away from 0. Thus, for every $f\in F_\varepsilon$, there exists $V_f\in \mathcal{V}$ such that
	$
	\inner{V_f}{f}>\inner{V}{f}.
	$
	By the continuity of the evaluation (\citet[Corollary 6.40]{aliprantis2007infinite}), for every $f\in F_\varepsilon$, there exists a weak*-open neighborhood $O_f$ of $f$ such that for all $f'\in O_f$, 
	$
	\inner{V_f}{f'}>\inner{V}{f'}.
	$
	Thus, $\{O_f:\: f\in F_\varepsilon\}$ is a weak*-open cover of $F_\varepsilon$. The functionals $f\in F_\varepsilon$ that expose a point in $\mathcal{V}$ are norm-dense in $ F_\varepsilon$ (see \citet{lau1976strongly} and note that $F_\varepsilon$ has non-empty interior in $L^\infty(\Theta)$).
	
	By compactness, we may then select a finite subcover $\{O_m:\:m=1,\ldots,M\}$ such that, for every $m=1,\ldots,M$, there exists $f'\in O_m$ such that $V_m:=V_{f'}\in\exp_+ \mathcal{V}$. Let
	$$
	G=\left\{\left(\inner{V_1-V}{f},\ldots,\inner{V_m-V}{f}\right)\mid f\in F_\varepsilon\right\}\subset\mathbb{R}^M.
	$$
	The set $G$ is compact and convex (because it is the image of the weak*-compact convex set $F_\varepsilon$ under a weak*-continuous linear map). Moreover, by construction of the open cover $\{O_m\}$, $G\cap \mathbb{R}^M_{-}=\emptyset$, where $\mathbb{R}^M_{-}$ the negative orthant.
	By the Separating Hyperplane theorem, there exists a vector $\alpha\in \mathbb{R}^M_+\setminus\{0\}$, such that $\alpha \cdot y> 0$ for all $y\in G$.  Renormalize $\sum_{i=1}^M \alpha_i=1$.
	
	Define 
	$
	\tilde V_\varepsilon=\sum_{i=1}^M \alpha_iV_i\in \mathcal{V}.
	$
	For all $f\in F_\varepsilon$,
	$$
	\inner{\tilde V_\varepsilon}{f}-\inner{V}{f}=\alpha\cdot \left(\inner{V_1-V}{f},\ldots,\inner{V_m-V}{f}\right)>0. 
	$$
	
	Now consider a sequence $\varepsilon_n\to 0$ and the corresponding sequence of $\tilde V_{\varepsilon_n}$ constructed above. Since $\mathcal{V}$ is norm-compact, a subsequence of $(\tilde V_{\varepsilon_n})$ converges to some $\tilde V\in \mathcal{V}$. 
	We show $\tilde V=V$, which proves the lemma. Recall that $V$ is undominated and suppose $\tilde V\neq V$. Then there exists a set $\tilde\Theta\subset\Theta$ of non-zero Lebesgue measure such that $V(\theta)>\tilde V(\theta)$ for all $\theta\in\tilde\Theta$. Thus, any density $f$ concentrated on $\tilde\Theta$ is such that 
	$
	\inner{V}{f}>\inner{\tilde V}{f}.
	$
	By norm-norm continuity of the evaluation, there exists a strictly positive density $f'$ and some $\tilde V_{\varepsilon_n}$ for $n$ large enough such that 
	$
	\inner{V}{f}>\inner{\tilde V_{\varepsilon_n}}{f'},
	$
	a contradiction.
\end{proof}

\begin{proof}[Proof of \Cref{lem:undominated_optimal}]
	Similar to the proof for \Cref{lem:undominated1}. Begin by assuming that $V$ is undominated but strictly suboptimal for all probability densities $f\in L^\infty(\Theta)$. Then, repeat the previous construction with $\varepsilon=0$ to obtain $\tilde V\in\mathcal{V}$ such that $\inner{\tilde V-V}{f}>0$ for all probability densities $f\in L^\infty(\Theta)$. This implies $\tilde V>V$ a.e., a contradiction.
\end{proof}

We extend \Cref{lem:undominated1} to cover all undominated mechanisms.
\begin{lemma}
	\label{lem:undominated2}
	$\und\mathcal{V}\subseteq \cconv (\exp_+\mathcal{V})$.
\end{lemma}

\begin{proof}
	Suppose not, that is, $V\in \und\mathcal{V}\setminus \cconv (\exp_+\mathcal{V})$. By \Cref{lem:undominated1}, $V\notin \underline{\und}\mathcal{V}$, that is, $V$ is optimal for some probability density $f^*\in L^\infty(\Theta)$ that is uniformly bounded away from 0.
	Since $\cconv (\exp_+\mathcal{V})$ is a closed subset of the norm-compact convex set $\mathcal{V}$, it is norm-compact. By the Hahn-Banach Separation theorem, there exists $f\in L^\infty(\Theta)$ such that
	$$
	\inner{V}{f}>\max_{V'\in \cconv (\exp_+\mathcal{V})} \inner{V'}{f}.
	$$
	Then $\tilde f=\varepsilon f +(1-\varepsilon) f^*$ (renormalized so that $\int \tilde f=1$) is a probability density for all sufficiently small $\varepsilon\in (0,1)$ since $f^*$ is uniformly bounded away from 0. Moreover,
	$
	\inner{V}{\tilde f}>\max_{V'\in \cconv (\exp_+\mathcal{V})} \inner{V'}{\tilde f}
	$
	by norm-norm continuity of the evaluation and Berge's maximum theorem (for the right-hand side). By the result due to \citet{lau1976strongly} used in \Cref{lem:undominated1}, there is another density $\hat f$ arbitrarily close to $\tilde f$ in $L^\infty(\Theta)$ and therefore also uniformly bounded away from $\varepsilon$ that exposes a point $\hat V\in \mathcal{V}$. Again by continuity and Berge's maximum theorem, 
	$
	\inner{V}{\hat f}>\max_{V'\in \cconv (\exp_+\mathcal{V})} \inner{V'}{\hat f}.
	$
	By definition, $\inner{\hat V}{\hat f}>\inner{ V}{\hat f}$. Thus, the point $\hat V$ exposed by $\hat f$ cannot be in $\exp_+ \mathcal{V}$, a contradiction.
\end{proof}

\begin{proof}[Proof of \Cref{lem:v_exposed}]
	The claim is a consequence of Milman's converse to the Krein-Milman theorem (see, e.g.,  \citet{klee1957extremal}, Theorem 1.1). The theorem implies that $\ext\cconv (\exp_+\mathcal{V})\subseteq \cl\exp_+\mathcal{V}$ since $\cconv (\exp_+\mathcal{V})$ is compact and convex.
	In particular, by \Cref{lem:undominated2}, every undominated extreme point of $\mathcal{V}$ must be in $\cconv (\exp_+\mathcal{V})$. But since $\cconv (\exp_+\mathcal{V})$ is a convex subset of $\mathcal{V}$, every undominated extreme point of $\mathcal{V}$ must also be in $\ext \cconv (\exp_+\mathcal{V})$ and therefore arbitrarily close to a point in $\exp_+\mathcal{V}$. 
\end{proof}

\end{appendix}

\printbibliography
\end{document}